\documentclass[aps,prd,amsmath,10pt,amssymb,twocolumn,showpacs,superscriptaddress,raggedbottom,preprintnumbers]{revtex4-1}

\usepackage{graphicx}
\usepackage{dcolumn}
\usepackage{bm}
\usepackage{hyperref}
\usepackage[utf8]{inputenc}
\usepackage{xcolor}

\newcolumntype{C}[1]{>{\centering\let\newline\\\arraybackslash\hspace{0pt}}m{#1}}

\graphicspath{{figures/}}

\begin{document}

\title{Understanding the energy resolution of liquid argon neutrino detectors}

\author{Alexander Friedland}
 \email{alexfr@slac.stanford.edu}
 
\author{Shirley Weishi Li}%
 \email{shirleyl@slac.stanford.edu}
\affiliation{SLAC National Accelerator Laboratory, 2575 Sand Hill Road, Menlo Park, CA, 94025}

\date{January 21, 2019}

\begin{abstract}
Available estimates for the energy resolution of DUNE vary by as much as a factor of four. To address this controversy, and to connect the resolution to the underlying physical processes, we build an independent simulation pipeline for neutrino events in liquid argon, combining the public tools {\tt GENIE} and {\tt FLUKA}. Using this pipeline, we first characterize the channels of non-hermeticity of DUNE, including subthreshold particles, charge recombination, and nuclear breakup. Particular attention is paid to the role of neutrons, which are responsible for a large fraction of missing energy in all channels. Next, we determine energy resolution, by quantifying event-to-event stochastic fluctuations in missing energy. This is done for several sets of assumptions about the reconstruction performance, including those available in the literature. The resulting migration matrices, connecting true and reconstructed neutrino energies, are presented. Finally, we quantify the impact of different improvements on the experimental performance. For example, we show that dropping particle identification information degrades the resolution by a factor of two, while omitting  charge deposits from de-excitation gammas worsens it by about 25\%. In the future, this framework can be used to assess the impact of cross section uncertainties on the oscillation sensitivity.
\end{abstract}
\preprint{SLAC-PUB-17352}

\maketitle


\section{Introduction}


In the years since the seminal discovery of the neutrino masses, neutrino physics has undergone significant evolution. In the early generations of experiments, the oscillation effects were \emph{large}, the observables {\it robust}, and the analyses could be done in simple two-flavor reductions of three-flavor mixing. For example, the observed deficit of the $^8$B solar neutrinos was as large as a factor of three. The solar neutrino problem was conclusively resolved when a robust observable, the ratio of charged- to neutral-current event rates, was measured by the SNO experiment~\cite{Aharmim:2011vm}. Finally, the results could be understood considering an effective two-state adiabatic level crossing in solar matter, with the mass-squared splitting of $\Delta m_\text{sol}^2 \simeq 7.4 \times 10^{-5}$ eV$^2$. Similarly, for atmospheric neutrinos the robust observable was the up/down asymmetry, its value was a factor of two, and simple $\nu_\mu\rightarrow\nu_\tau$ oscillations with $\Delta m_\text{atm}^2 \simeq 2.4 \times 10^{-3}$ eV$^2$ could explain the data~\cite{Fukuda:1998mi}.

By comparison, modern neutrino oscillation experiments target subtle 3-flavor effects, such as CP violation and the signatures of the different mass hierarchies. They typically require the oscillation probabilities to be measured with an accuracy of 10\% or better. 
Possible Beyond-the-Standard-Model (BSM) physics effects could further complicate the phenomena. As the field enters its precision era, with NO$\nu$A~\cite{NOvA:2018gge} and T2K~\cite{Abe:2011ks} collecting data and the Deep Underground Neutrino Experiment (DUNE) in the design stage~\cite{Acciarri:2015uup}, future experimental success \emph{requires} accurate modeling and error estimation.

One key quantity that needs to be measured precisely is the neutrino energy, $E_\nu$. This is obvious already from the fact that the mass-squared splittings between the vacuum Hamiltonian eigenstates, $\Delta m_{i}^2$, enter the oscillation probabilities in combinations $\Delta m_{i}^2/E_\nu$. The locations of features in the oscillated energy spectrum, therefore, provide direct information on $\Delta m_{i}^2$, and mis-calibration of the energy scale can lead to a mis-measurement of $\Delta m_{i}^2$.  In turn, resolution effects smear the signal over neutrino energies, washing out the oscillation features. Mis-modeling of the amount of this smearing can be mistaken for incorrect values of the mixing angles $\theta_{i}$. 

Moreover, since neutrino cross sections depend on $E_\nu$, energy mis-reconstruction can lead one to  incorrectly infer the appearance probability in the $\nu_\mu\rightarrow\nu_e$ or $\bar\nu_\mu\rightarrow\bar\nu_e$ modes. This can spoil the measurements of the CP-violating phase, $\delta_\text{CP}$, or introduce apparent contradictions between different pieces of data.

Even more impetus for accurate energy determination arises if one relaxes the no-BSM-physics assumption in the oscillation analysis. With possible nonstandard interactions (NSI), one finds that, at a given energy, the same appearance outcomes, for both neutrinos and antineutrinos, can be reproduced for several different parameter values (as discussed, e.g., in the context of NO$\nu$A in Ref.~\cite{Friedland:2012tq}). This degeneracy, however, is broken if one measures the appearance probability over a range of energies.


The performance of the calorimetric method is tied to the \emph{hermeticity} of the detector. Both average missing energy and its event-to-event fluctuations must be accurately modeled. The former is essential for a bias-free reconstruction of the neutrino energy scale. The latter lead to finite energy resolution~\cite{LBNO}, as discussed in detail later.  Thus, missing energy and energy resolution are inextricably linked. To understand both, a systematic study of all relevant energy-loss modes is required.
%

The fraction of the energy that is missed depends on the properties of the hadronic system produced in neutrino interactions. To accurately predict these properties is a highly nontrivial task at energies of a few GeV, where neutrino interactions are described by neither low-energy nuclear physics methods, nor perturbative QCD applicable to the deep inelastic scattering (DIS) regime. 
One may wonder why DUNE  has chosen such a challenging energy range. The answer comes from the physics of three-flavor oscillations~\cite{Diwan:2004bt}. To distinguish the mass hierarchies, the experiments take advantage of the matter effect in the Earth's crust. The matter term in the oscillation Hamiltonian, $\sqrt{2} G_F n_e$, has the dimension of inverse distance and the magnitude $\sim (2 \times 10^3$ km$)^{-1}$. In other words, baselines of $\mathcal{O}(10^3)$ km are required to have significant matter effects. Indeed, DUNE will have a baseline of 1,300 km. But then, given the value of the atmospheric mass-squared splitting, $\Delta m_\text{atm}^2$, one is led to energies of 2--3 GeV to have the first oscillation minimum at the far detector. 

Our investigation connects to several previous studies. 
Two early papers have focused on missing energy in the primary neutrino interaction: one studied prompt neutrons~\cite{Ankowski:2015kya} and another prompt particle thresholds~\cite{Ankowski:2015jya}. Since these studies do not supply explicit information about fluctuations in these channels, we cannot assess the impact of these energy-loss channels on energy resolution. Besides, as we will see, to fully quantify missing energy, it is important to also model the subsequent particle propagation in medium.

As for the energy resolution, the approach used in Ref.~\cite{Ankowski:2015jya}, as well as in the DUNE Conceptual Design Report (CDR)~\cite{Acciarri:2015uup}, is to apply a set of thresholds and model the effects of the full propagation with effective energy loss and Gaussian smearing prescriptions applied to particles produced at the primary vertex. The simulation in Ref.~\cite{Acciarri:2015uup} is carried out with the so-called Fast Monte Carlo (FastMC) code, with rather conservative particle threshold values. The resulting migration matrices~\cite{Alion:2016uaj} form the basis of most modern oscillation forecasts for DUNE. Reference~\cite{Ankowski:2015jya} obtains an alternative set of migration matrices, reflecting different prescriptions applied to prompt particles.

A qualitatively different method was put forward in Refs.~\cite{Sorel:2014rka, DeRomeri:2016qwo}. In those papers, it was suggested to sum up the ionization charges from all particles in the hadronic shower, with no thresholds. Events were simulated using the {\tt LArSoft} package and cosmogenic and radiogenic backgrounds were neglected. The resolution of the total-charge method for contained events was found to be $\sim$5\%~\cite{Sorel:2014rka}, some factor of four better than in the CDR document. The same general method is followed in the recent reconstruction studies~\cite{Grant2018DPFtalk}, which, however, arrive at different (lower) resolution, despite improving the procedure by considering the lepton shower and the hadronic system separately. The alternative migration matrices released in Ref.~\cite{DeRomeri:2016qwo} differ markedly from those in Refs.~~\cite{Acciarri:2015uup, Alion:2016uaj}. It is also puzzling, on general physical grounds, that one is able to achieve such good resolution despite discarding all information about the event composition. What would the resolution be had the information been kept? 

\begin{figure}[t]
	\begin{center}
        \includegraphics[width=0.8\columnwidth]{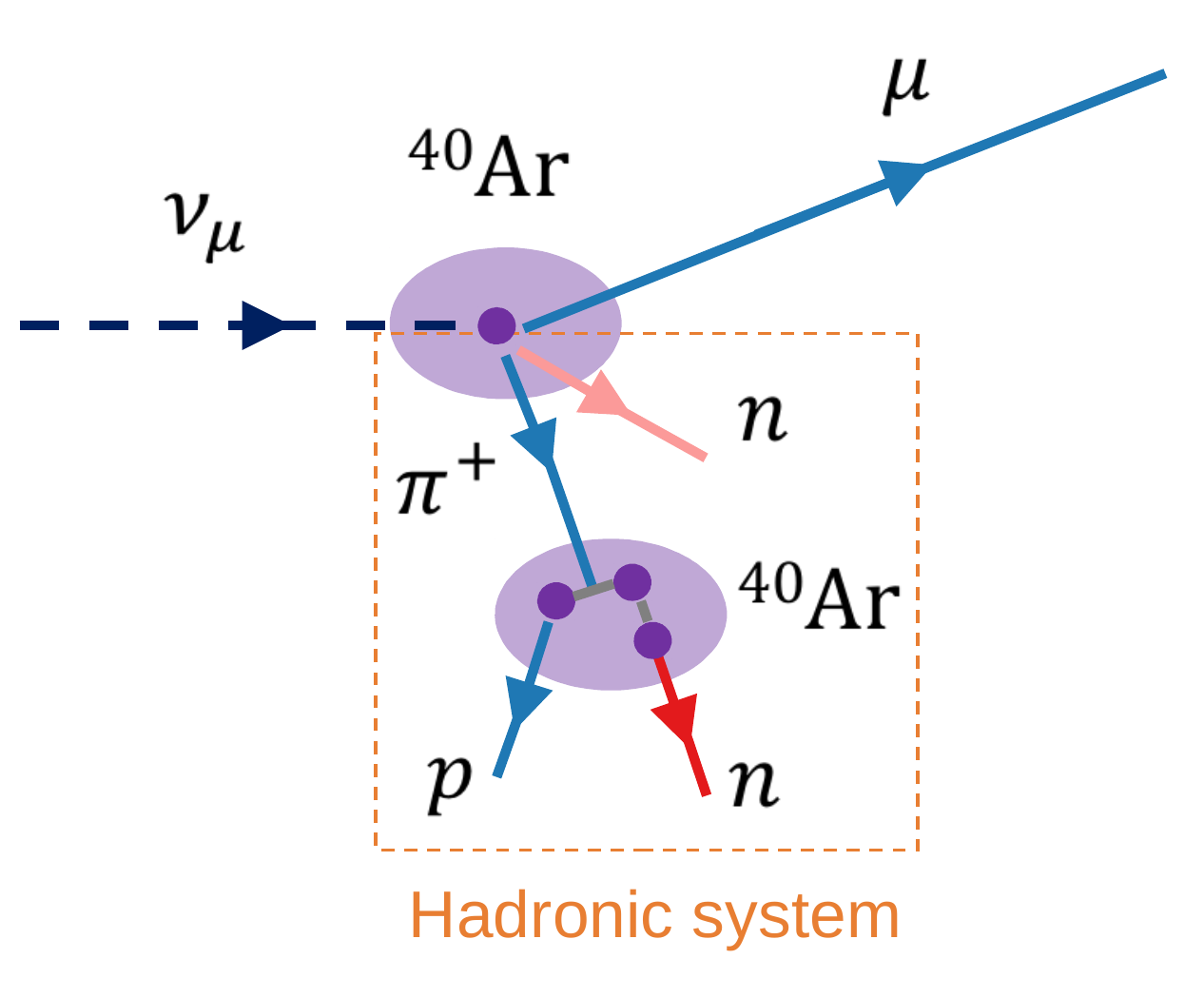}
        \caption{A neutrino event at DUNE: a conceptual illustration.}
        \label{fig:neutrino_cartoon}
    \end{center}
\end{figure}

Our aim is to clarify this situation, but also to characterize the role of different physical processes in energy loss and resolution at DUNE. The latter may be not easy to tease out from a full detector simulation. What we would like here is to separate effects that can be remedied---too-small simulation volume, too-high energy thresholds, information about the events that can be kept---from the limitations that are intrinsic to the detection process in liquid argon. Accordingly, our study is, in a way, deliberately schematic. We stress that our work is not a substitute for detailed detector simulations---with wires, electronic noise, and cosmogenic and radiogenic backgrounds---or for event-reconstruction studies. Rather, we aim to encourage further such studies by our findings. 


\section{Motivations and presentation outline}

To set up the problem, let us consider a typical charged-current (CC) neutrino interaction event in DUNE. The primary interaction has the form 
\begin{equation}
    \nu_l + A \rightarrow l + \text{hadrons} + X.
    \label{eq:nu_int}
\end{equation}
The outgoing lepton $l$ is either an electron or a muon \footnote{The interaction for $\nu_\tau$ has a high threshold, $\sim 3$~GeV~\cite{Formaggio:2013kya}, so its rate is much smaller.}; $X$ is the remnant nucleus.  All other outgoing particles are conventionally called ``hadrons,'' even though this system, in addition to protons, neutrons, charged pions and kaons, could also contain gamma rays (for example from $\pi^0$ decays; see later).

Since the DUNE beam neutrinos have energies of 1--4 GeV, the hadronic system can have a rich and varied structure.  Figure~\ref{fig:neutrino_cartoon} depicts, schematically, a possible reaction:  the neutrino interaction at the primary vertex creates a hadronic resonance, which decays to a charged pion and a neutron. Additional hadrons could be knocked out of the primary nucleus due to final-state interactions (not shown).
All these prompt particles then propagate through the detector and---in addition to ionization---can cause secondary interactions, knocking out extra nucleons, as well as creating pions and $\gamma$ rays. Bremsstrahlung radiation and nuclear de-excitations produce additional, low-energy $\gamma$'s. 
To relate the resulting ionization charge to the neutrino energy, full modeling of the propagation process is required. 

Even before running the full simulations, however, it should be obvious that not all of the original neutrino energy ends up in detectable ionization charge. Let us consider some examples. First, the propagation process increases particle multiplicity and reduces their average energies. As the resulting cascade fully develops, some particles become difficult to detect. One is therefore naturally led to the concept of detection thresholds as one of the ways energy can be missed. Second, propagating hadrons can disrupt a number of argon nuclei in the medium. The energy spent on this nuclear breakup does not all translate to ionization. Third, some energy goes to neutrinos in pion and muon decays, which escape the detector. Our first task is to quantify the contributions of these and other energy loss channels to the overall energy flow in DUNE events.

\begin{figure}[t]
	\begin{center}
        \includegraphics[width=\columnwidth]{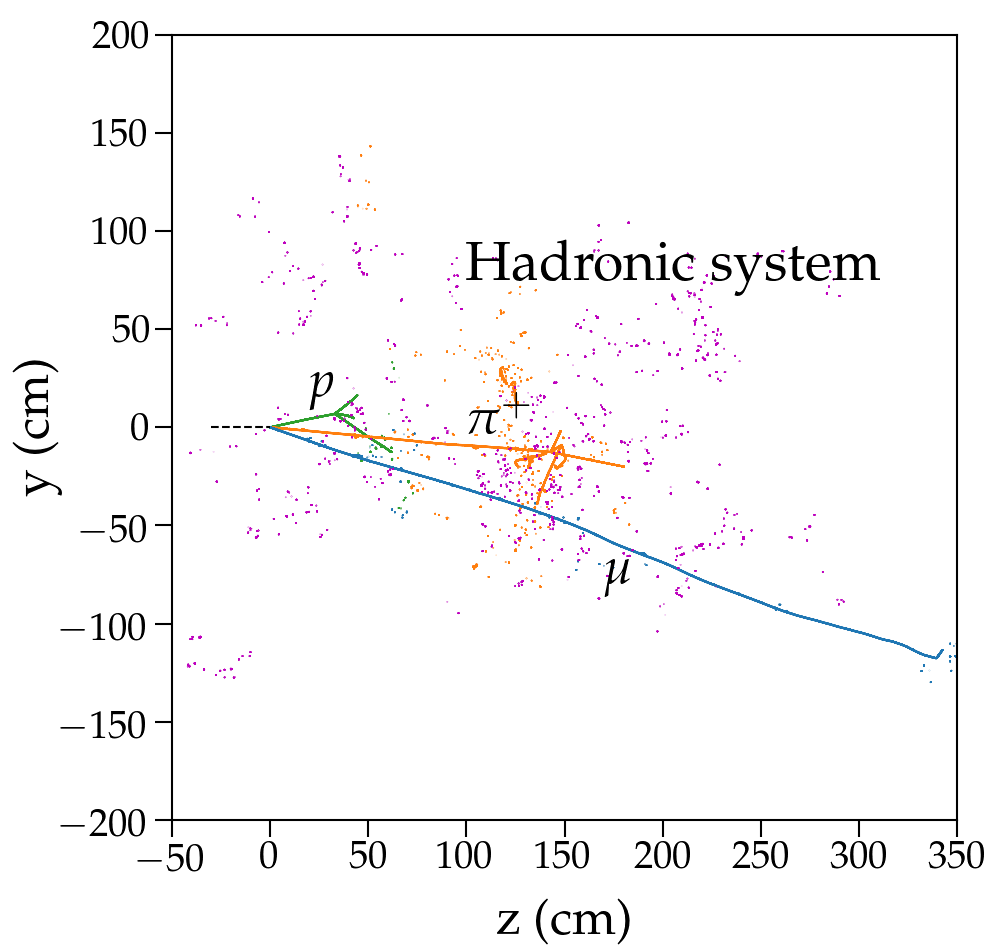}
        \caption{An example simulated 4 GeV $\nu_\mu$ event using {\tt GENIE} and {\tt FLUKA}. The magenta energy deposits are caused by neutrons undergoing multiple scatterings; the orange color denotes energy originally carried by the prompt charged pion.}
        \label{fig:neutrino_event}
    \end{center}
\end{figure}

Figure~\ref{fig:neutrino_event} shows an actual event from our simulations, in which a muon, a $\pi^+$, a proton, and two neutrons are exiting the primary vertex. All the phenomena outlined above are present. The charged hadrons are seen to undergo secondary interactions, creating additional tracks. The neutrons, being neutral, themselves do not leave ionization tracks and can only be seen through charged particles created in secondary interactions. Their energy is dissipated via numerous subthreshold particles and nuclear breakup. As will be seen later, they can also create secondary hadronic showers, and these can be meters away from the primary interaction. Neutrons thus present a special challenge and we designate them in a special category. 

Notice that these considerations apply to both prompt and secondary particles; to quantify the importance of each missing energy channel one has to model the entire event. Accordingly, we built a framework which combines a neutrino event generator, {\tt GENIE}, with a propagation code, {\tt FLUKA}. Using this framework, we model neutrino and antineutrino interactions inside a liquid argon detector and simulate a large number of scattering events for the energies relevant to DUNE. 

The presentation is organized as follows. We begin, in Sec.~\ref{sec:events}, by reviewing the processes occurring in the primary interaction vertex and describing the prompt particles that can be created. We then describe how each of these particle types propagate through the detector medium, liquid argon. This part is essential for understanding the physics behind our findings. However, it can be skipped at first reading by readers primarily interested in our simulation results.

After this introduction, in Sec.~\ref{sect:example10}, we discuss a small set of our simulated events, which will be seen to have both sizable average missing energy and large event-to-event variations. This motivated our two main analysis goals.

The first goal is to establish the average contribution of each missing energy channel. This question is answered in Sec.~\ref{sect:averagelosses} as a function of neutrino energy, for both neutrino and antineutrino scattering. This establishes the \emph{average} conversion functions between visible charge and the true neutrino energy. Our second goal is to characterize the event-by-event \emph{dispersion} in the visible charge. This dispersion leads to an intrinsic limitation on how well the hadronic energy of each event can be reconstructed, i.e., to finite energy resolution. We report, in Sec.~\ref{sect:modelresolution}, the resolution numbers of our simulations, under different sets of assumptions about reconstruction performance. This procedure yields a set of \emph{migration matrices}, connecting visible and true hadronic energies, which can be used as inputs to oscillation studies.  The implications of these results are further studied in Sec.~\ref{sect:averagelosses}. We compare the impact of different improvements on the energy resolution, which can inform experimental priorities. We also categorize energy loss channels in those that can be improved and those that are intrinsic to this detector technology. Broader implications of our results and outlook for the future are presented in Sec.~\ref{sect:conclusions}.


\section{Modeling neutrino events}
\label{sec:events}

Neutrino-nucleus scattering and subsequent propagation of daughter particles in the medium are  complex physical processes requiring dedicated, extensive simulation codes. Our simulation framework in this paper is built on two community-based packages: we use {\tt GENIE} (version 2.12.8)~\cite{Andreopoulos:2009rq} for primary neutrino interactions and {\tt FLUKA} (version 2011.2x.2)~\cite{Bohlen:2014buj, Ferrari:2005zk} for particle propagation in liquid argon. {\tt GENIE} is the default generator code used in DUNE studies (as well as all other Fermilab-based neutrino experiments), while {\tt FLUKA} is a well-tested package for calculations of particle transport and interactions with matter, which has been proven to provide good description of MeV hadronic physics~\cite{Acciarri:2018myr}. 

While not a substitute for full detector simulations, event reconstruction and analysis capabilities of the {\tt LArSoft} software package \cite{Church:2013hea,larsofturl}, our framework offers a number of complementary features that suit our present purpose. The chief among these are speed, flexibility and transparency to underlying physics assumptions.  We also note that we employ no internal or proprietary software or configuration files from DUNE. All results presented here should thus be fully reproducible using only publicly available resources.


\subsection{Primary neutrino-nucleus interaction}
\label{sect:primary_interactions}

The {\tt GENIE}~\cite{Andreopoulos:2009rq} package simulates primary neutrino interactions, $\nu_l \,+ \,^{40}$Ar $\rightarrow l + \text{hadrons} + X$.  Our focus is on the CC process, for which the final-state lepton $l$ is charged and leaves either a clear ionization track (when it is a muon) or an electromagnetic shower (when it is an electron). Tau lepton production can be neglected for the DUNE beam. For every event, {\tt GENIE} generates the composition of the full final state and returns the four-momenta for all outgoing particles.  

\begin{figure}[t]
	\begin{center}
        \includegraphics[width=\columnwidth]{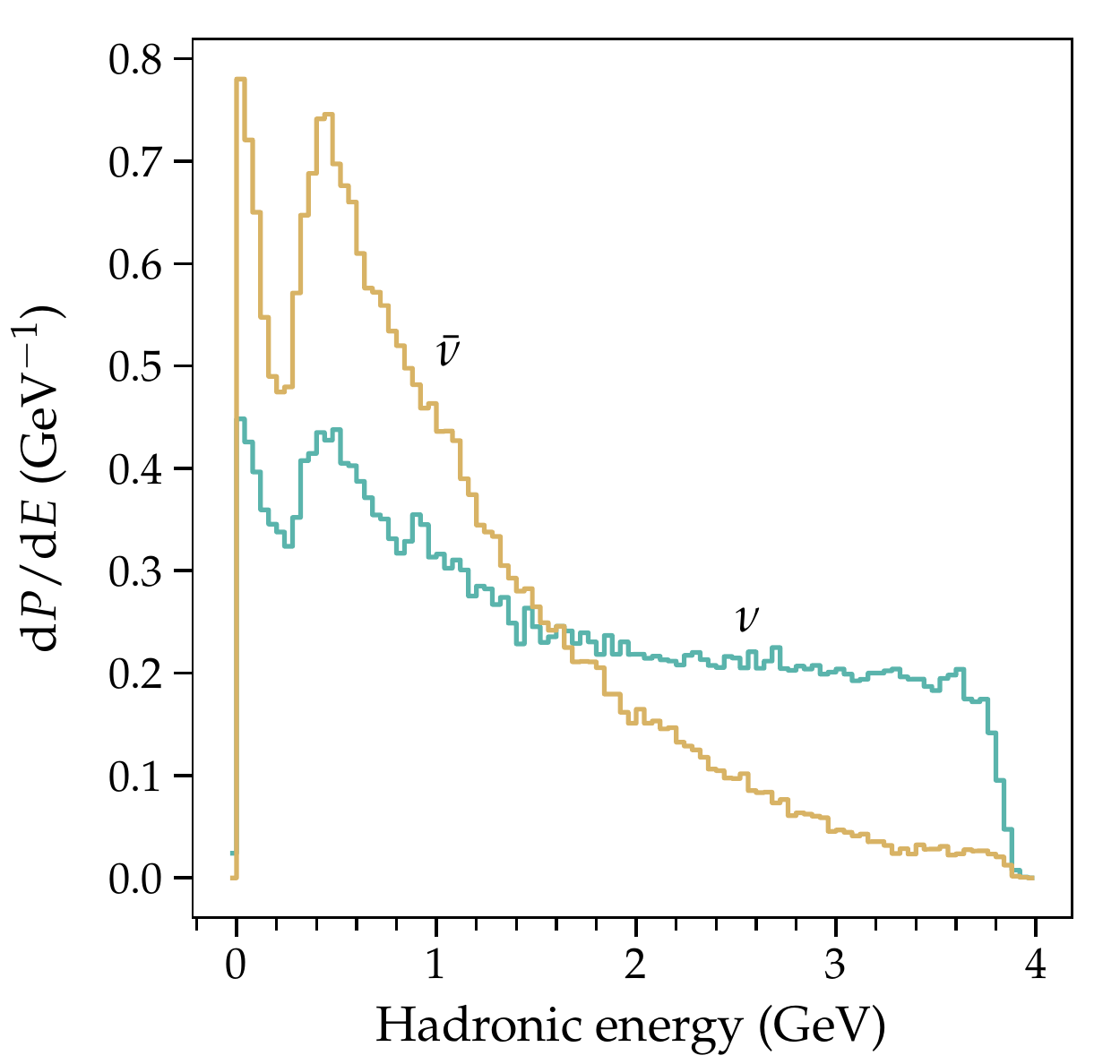}
        \caption{Hadronic energy distributions produced in the scattering on argon of 4 GeV neutrinos (blue) and antineutrinos (brown). The average hadronic energy is 1.6 GeV for $\nu$, and 1.0 GeV for $\bar{\nu}$.}
        \label{fig:primary_hadronic}
    \end{center}
\end{figure}

The range of neutrino energies at DUNE is 1--4 GeV and our knowledge of the relevant interaction physics in this regime is uncertain. Unlike in the case of neutrino-electron scattering, no closed analytical expressions are available for neutrino scattering on hadrons. The generator codes by necessity invoke approximate models for several scattering regimes, along with empirical prescriptions combining, and interpolating between, these regimes. Several such models are implemented in {\tt GENIE} and the results, strictly speaking, depend on which settings (or ``tunes") of the code are used.  Moreover, it is established that there are differences among the different generators, such as {\tt GENIE}, {\tt NuWro}~\cite{nuwrourl}, or {\tt GiBUU}~\cite{Buss:2011mx,gibuuurl}, and between all generators and neutrino data~\cite{Acciarri:2015uup, Mosel:2016cwa}. This applies to both inclusive event rates and to the composition of final states, both of which are important for the performance of long-baseline experiments. 

The physics of neutrino-nucleus interactions remains an active area of research. A detailed investigation of the generator dependence falls outside the scope of this paper. We simply note that our results are obtained for a specific version of {\tt GENIE}, 2.12.8, with pre-compiled default cross sections (``\textsf{DefaultPlusMECWithNC}"), and will change with different choices.

The basic characteristic of $\nu$N CC interactions is the fraction of the initial neutrino energy, $E_{\nu}$ that goes into the final-state hadronic system, $E_\text{had}$. This fraction, $y \equiv E_\text{had} / E_\nu$, is traditionally called ``inelasticity."  Let us consider Fig.~\ref{fig:primary_hadronic} and, for the moment, focus on the blue histogram, which shows the distribution of $E_\text{had}$ produced by a 4 GeV muon or electron neutrino. Observe that this distribution is broad, ranging from $E_\text{had}\sim0$ to $E_\text{had}\sim E_{\nu}$.  
%
At the qualitative level, this behavior can be understood as a consequence of (1) the contact nature of the weak interactions at these energies and (2) the helicity structure of the interaction. 

Concerning the first point, it is instructive to draw a contrast with what happens in electron-nucleus scattering. The latter is strongly forward-peaked, thanks to the masslessness of the photon, so that most of the energy in a typical scattering event remains in the electron. In contrast, neutrino scattering does not suffer from the forward Rutherford  divergence of the Coulomb scattering cross section.

\begin{figure}[t]
	\begin{center}
        \includegraphics[width=\columnwidth]{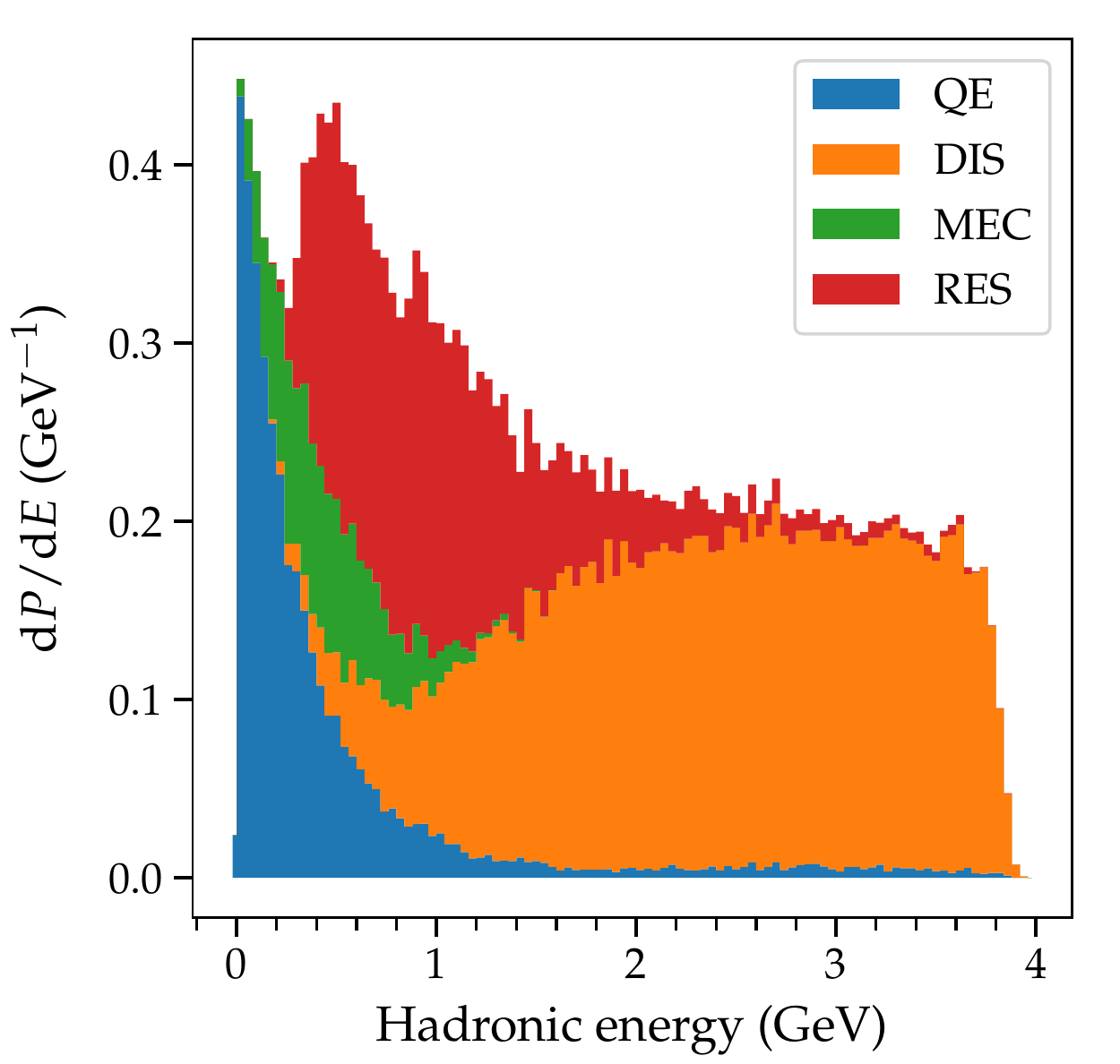}
        \caption{Same as the blue histogram in Fig.~\protect{\ref{fig:primary_hadronic}}, but broken down according to the physical processes involved (per {\tt GENIE} default tune).}
        \label{fig:primary_breakdown_nu}
    \end{center}
\end{figure}

For the second argument, let us for a moment make a qualitative approximation of the CC $\nu$N interaction as DIS on the constituent  quarks, followed by subsequent hadronization. In the neutrino-quark center of mass (CM) system, the angular distribution of the scattering products is determined by helicity arguments. Since only left-handed fields participate in CC weak interactions, the neutrino-quark initial state has an angular momentum of 0, while the neutrino-antiquark system has an angular momentum of 1. Consequently, in the first case the angular distribution in the CM frame is isotropic, while in the second case it is proportional to $|d_{11}^{(1)}(\Theta)|^2=|(1+\cos\Theta)/2|^2$, where $d_{m'm}^{(j)}(\Theta)=\langle j m'| \exp(-i\Theta J_y)|jm\rangle$ is the usual Wigner $d$-matrix for angular momentum and $\Theta$ is the CM scattering angle. The distribution in $\Theta$ can then be related to the distribution in the Mandelstam invariant $t$, using $t=(p_\nu-p_l)^2\simeq - E_\nu x m_N (1-\cos\Theta)$, where $m_N$ is the nucleon mass. The invariant $t$, in turn, can be related to the distribution of the lepton energy loss $E_\nu-E_l=-t/(2 x m_N)$ using simple kinematics ($x$ is the momentum fraction carried by the quark, which is treated as massless). The net result is that the $\nu q$ CC cross section is flat in $y$, while  $\nu\bar{q}$ has a $(1-y)^2$ dependence. As the blue histogram in Fig.~\ref{fig:primary_hadronic} shows, the quark parton distribution function dominates the nucleons, with sea antiquarks providing a small increase at lower $y$ (at lower $E_{\rm had}$). 

The situation is reversed for $\bar\nu$N CC scattering. In this case, the helicity arguments give that the $\bar\nu q$ cross section has a $(1-y)^2$ dependence, while the $\bar\nu \bar q$ cross section is flat. This results in a significantly softer distribution of the hadronic products, as illustrated by the brown histogram in Fig.~\ref{fig:primary_hadronic}. While on average 40\% of the neutrino energy goes into the hadronic system in $\nu$N CC scattering, in the case of $\bar\nu$N CC scattering the average hadronic energy fraction is only 25\%.

Of course, at low values of $E_{\rm had}$ it is not appropriate to treat the scattering process in the DIS approach. Figure~\ref{fig:primary_hadronic} indeed shows clear deviations from the DIS behavior at $E_{\rm had} < 1$ GeV, where peaks are seen. The lowest energy peak corresponds to the CC quasi-elastic (CCQE) channel, $\nu_{\mu}+n\rightarrow\mu+p$. The second peak is due to the interaction channel forming the $\Delta$ resonance, $\nu_\mu + n \rightarrow \mu+\Delta$.  There are a few higher resonances that are not quite visible in the figure and these are modeled in {\tt GENIE} to gradually transition into the DIS regime. In the region between the CCQE peak and the $\Delta$ resonance, the cross section also receives contributions from multinucleon effects (the so-called meson-exchange currents). All these processes are illustrated in Fig.~\ref{fig:primary_breakdown_nu}, which shows the breakdown of the blue histogram in Fig.~\ref{fig:primary_hadronic} into its physical components, as given by the default {\tt GENIE} tune. 

Finally, hadrons experience final-state interactions on the way out of the argon nucleus. This creates an intranuclear cascade, which can leave the nucleus in an excited state, absorb pions and knock out further nucleons.

The resulting hadronic system is complex and its composition is varied and energy dependent.  One neutrino interaction can typically produce 2--5 hadrons, which can be protons, neutrons, charged pions, or gammas created in the decays of neutral pions. Nuclear de-excitations can also yield gammas, with lower energies. 

\begin{figure}[t]
	\begin{center}
        \includegraphics[width=\columnwidth]{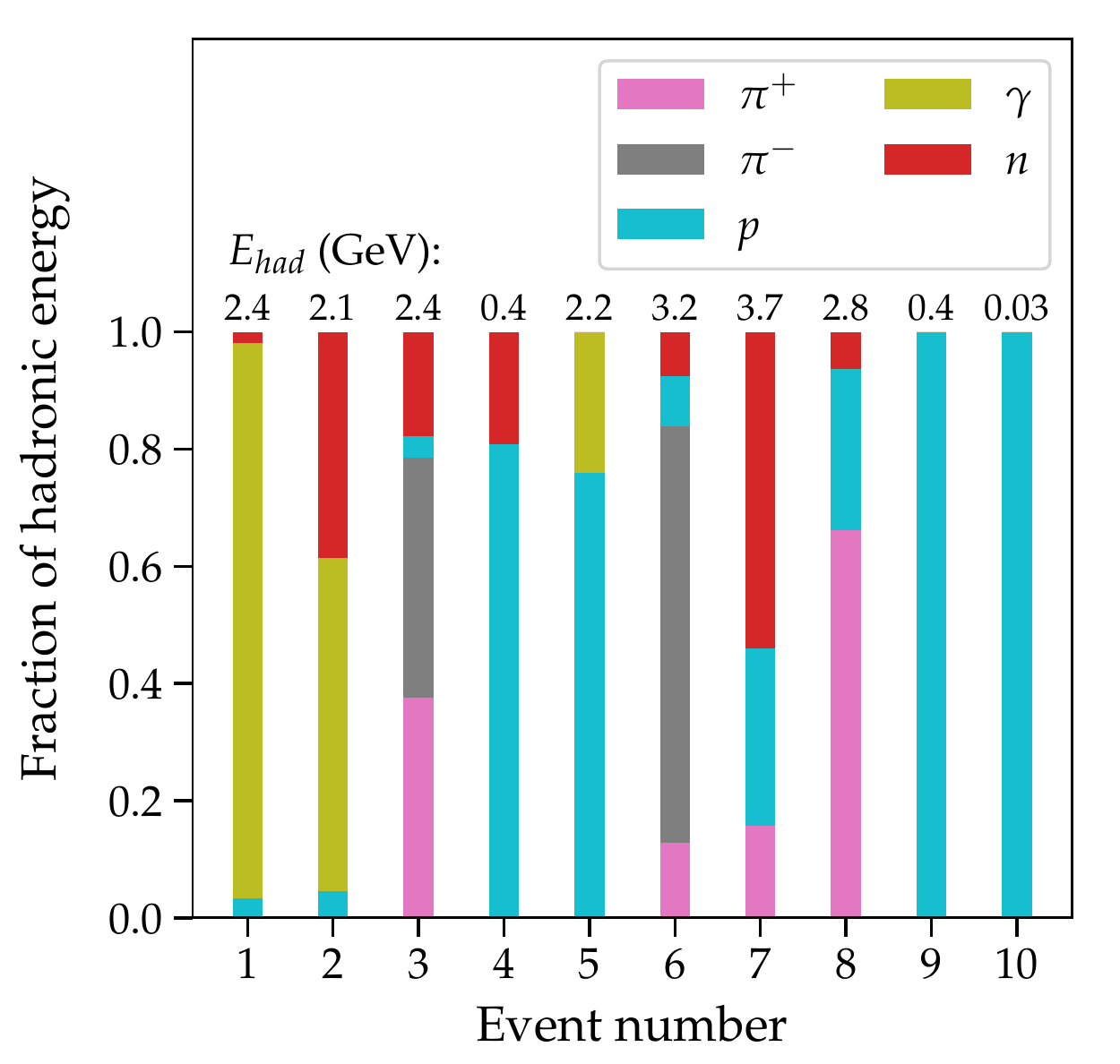}
        \caption{Hadronic composition of 10 CC $\nu_\mu+^{40}$Ar scattering events. The neutrino energy is 4 GeV in each event.}
        \label{fig:hadronic_fluctuation}
    \end{center}
\end{figure}

A helpful illustration is provided by Fig.~\ref{fig:hadronic_fluctuation}, which shows the total energy and the composition of the final-state hadronic system for a set of ten CC $\nu_\mu+^{40}$Ar scattering events generated by {\tt GENIE}. The neutrino energy in each case is 4 GeV. We can easily see that there are large fluctuations in the energy of the hadronic system, $E_\text{had}$: it varies from nearly all available energy (event 7) to nearly zero (event 10). Also apparent are large variations in the composition of this system.  Compare, for example, events 1 and 3, which have the same value of $E_\text{had}$, 2.4~GeV: in one case, this energy comes out as an electromagnetic shower, while in the other, as charged hadrons and neutrons. 

\begin{figure}[t]
	\begin{center}
        \includegraphics[width=\columnwidth]{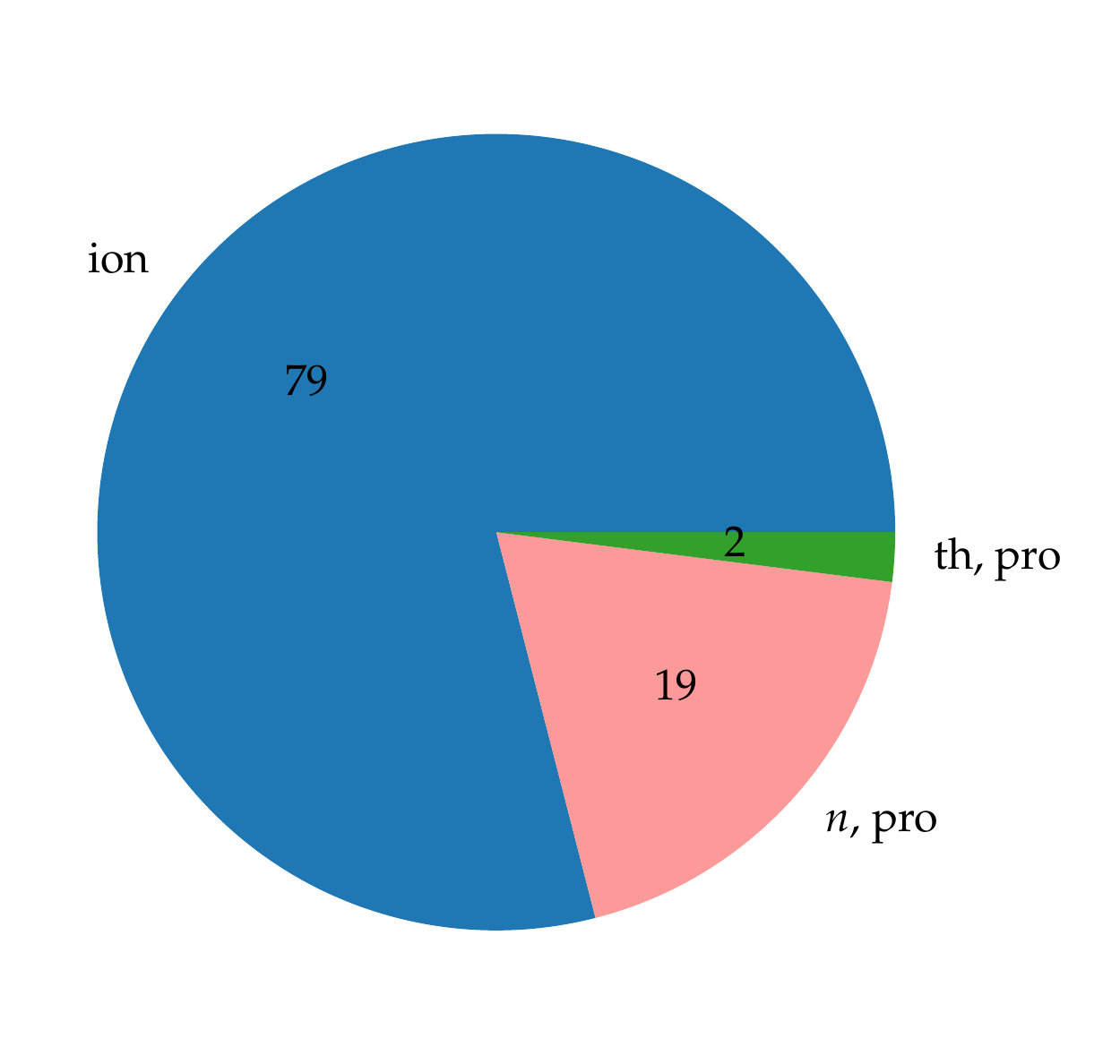}
        \caption{Hadronic energy budget after primary neutrino interaction.  A set of 10,000 4 GeV $\nu_\mu+^{40}$Ar scattering events has been averaged over. Shown are the fractions of the hadronic energy that go into prompt neutrons (\emph{n, pro}), subthreshold particles according to Table~\ref{tab:threshold} (\emph{th, pro}) and the rest (\emph{ion}).}
        \label{fig:hadronic_budget_primary}
    \end{center}
\end{figure}

We also note, for completeness, that event 8 here has already been presented in Fig.~\ref{fig:neutrino_event}. The muon has a starting energy of 1.1 GeV and the hadronic system is dominated by the charged pion, with the proton also contributing.

The average energy fractions carried by different final-state hadrons are presented in Appendix~\ref{sect:Appendix_average_composition}. Notice that at neutrino energies of 3--4 GeV the average hadronic energy becomes approximately equipartitioned between the different hadron types: $p$, $n$, $\pi^+$ and $\pi^0$ for neutrinos and $p$, $n$, $\pi^-$ and $\pi^0$ for antineutrinos. Thus, at DUNE, all hadrons must be well understood.   

At this stage, one might be tempted to conjecture what the sources of missing energy are. The immediate candidate is neutrons, which do not leave charged tracks. The second is particle thresholds and detection efficiencies. This part depends on the performance of the data acquisition system and the event reconstruction software. As a conservative reference point, one could consider the threshold values quoted in the DUNE CDR document. These are reproduced in Table~\ref{tab:threshold}.  

\begin{table}[h]
    \caption{Detection thresholds according to the DUNE CDR document~\cite{Acciarri:2015uup}. The values given correspond to the kinetic energy of each particle.}
    \begin{center}
    \setlength{\extrarowheight}{1.5pt}
    \begin{tabular}{ | C{1.8cm} |C{.8cm} |C{0.8cm} |C{0.8cm} | C{0.8cm}|C{0.8cm}|C{1cm}|}
    \hline
    & $p$ & $\pi^\pm$ & $\gamma$ & $\mu$ & $e$ & others \\ 
     \hline
    Thresholds (MeV) & 50 & 100 & 30 & 30 & 30 & 50\\ 
    \hline
    \end{tabular}
    \label{tab:threshold}
    \end{center}
\end{table}

To quantify the impact of prompt neutrons and thresholds, we generated a set of 10,000 $\nu_\mu+^{40}$Ar CC scattering events with a 4 GeV neutrino energy and have taken the average. The results are shown in Fig.~\ref{fig:hadronic_budget_primary}.  Most energy is visible, i.e., carried by charged particles above thresholds.  There is 19\% energy loss to neutrons, a number that is consistent with Ref.~\cite{Ankowski:2015kya}.  Thresholds, on the other hand, play a negligible role here.  This may be a little surprising, as an intuitive argument is that if a charged pion falls below threshold, then we lose not only its kinetic energy, but also its rest mass of 139 MeV. In fact, pions on average carry $\sim$30\% of total hadronic energy.  But they most likely have around 500 MeV of the total energy, with a long tail extending up to 3 GeV. The fraction of pions that falls below the 50 MeV threshold is tiny.

We emphasize that Fig.~\ref{fig:hadronic_budget_primary} does not give the complete picture of energy loss in liquid argon neutrino detectors, because it takes into account only the primary neutrino interaction. For example, neutrons can be produced also downstream, as the events develop. Conversely, while neutrons themselves do not leave tracks, some of their energy can nonetheless be converted to visible charge, via hadronic interactions with the argon nuclei in the detector medium. It is also intuitively clear that most subthreshold particles will be found at the last stages of shower development. These arguments make it apparent that a meaningful study of the energy loss channels \emph{must} include the full event development.  We therefore turn to it next.


\subsection{Particle propagation in medium}
\label{sect:propagation}

We inject all final-states particles out of {\tt GENIE} into {\tt FLUKA}, with their correct 4-momenta. {\tt FLUKA} uses these inputs to simulate the full event development in liquid argon, incorporating all relevant physics processes, such as ionization and radiative energy losses, hadronic inelastic interaction, and particle decays.  Unlike {\tt GEANT4}~\cite{Agostinelli:2002hh, Allison:2006ve}, which handles particle propagation in {\tt LArSoft}, physics models in {\tt FLUKA} are not tunable by users.  

For each event, primary particles and all subsequently produced secondary particles interact and propagate until all particles either fall below propagation thresholds or escape a user-defined geometry. We set  our propagation thresholds to 0.05 MeV, which is much lower than the DUNE detection thresholds. As for the geometry settings, we define the interaction region to be 12~m $\times$ 14.5~m $\times$ 58~m, the geometry of one 10-kton DUNE module~\cite{Acciarri:2016ooe}.  When assuming neutrons to be 100\% invisible, we discard neutron propagation with the {\tt DISCARD} card.  

Different types of final-state particles have distinct signatures in liquid argon. Below we review what happens to muons, electrons, gamma rays, charged pions, protons, and neutrons. The latter deserve a special discussion, as they are a major channel of missing energy.


\subsubsection{Charged leptons and gamma rays}

\textit{Muons}. Charged particles, when moving through liquid argon, impart some of their kinetic energy to surrounding electrons. This results in ionization tracks. In a typical $\nu_\mu+^{40}$Ar scattering event, the longest track is left by a muon, as illustrated in Fig.~\ref{fig:neutrino_event}. This track is mostly straight, with some deflection provided by multiple Coulomb scattering on ions in the medium. The energy loss rate is almost constant, $\sim$2 MeV/cm~\cite{PDG}, as is expected from a minimally ionizing particle.  The distance a muon travels before coming to a stop can therefore be estimated by
\begin{equation}
    d_\mu = \frac{E_\mu}{0.2~\text{GeV/m}} .
\end{equation}
A 4 GeV muon can travel $\sim$20~m.  Each module of the far detector of DUNE is 58~m long and 12 m $\times$ 14.5~m in transverse measurements~\cite{Acciarri:2016ooe}, so most muons are contained. As an example, the containment fraction is 2/3 at $E_{\nu}=3$ GeV. Of the exiting muons, 2/3 do so through the sides and only 1/3 through the back of the detector. 

Even with a 20-meter-long track, a relativistic muon comes to rest faster than its decay lifetime (which gets further affected by time dilation). Therefore, muon energy loss due to decay in flight can be neglected. At the end of the muon trajectory in Fig.~\ref{fig:neutrino_event} one can notice a short track attached to it. This is a Michel electron from the decay $\mu\rightarrow e+\bar\nu_e+\nu_\mu$. 

The energy of contained muons can then be inferred from the total distance they travel, with a $\sim$5\% energy resolution~\cite{Grant2018DPFtalk}, or even more accurately from the total ionization charge, as discussed later. When a muon is only partially contained, its energy can be estimated by the rate of multiple scattering along the contained segment, i.e., by looking at deviations from a straight line.  This degrades energy resolution, with estimates ranging from 18\%~\cite{Acciarri:2016ooe} to  30\%~\cite{Acciarri:2015uup}. Notice that some of the existing simulations of muons at DUNE have been carried out in a reduced geometry, leading to a higher escape fraction and hence worse energy resolution for a ``typical'' muon. 

\textit{Electrons}. At DUNE, $\nu_e+^{40}$Ar
scattering produces final-state electrons in the GeV energy range, which create electromagnetic showers~\cite{PDG, Li:2015kpa}.  The shower develops as the electron emits gamma rays by bremsstrahlung, which in turn create more electrons (and positrons) by pair production.  With each generation in the cascade, particle multiplicity increases, while average energy decreases.  

The distance from the beginning to the maximum of the shower is approximately
\begin{equation}
     l_{max} \simeq X_{0}  \log(E / E_c).
\end{equation}
Here, $X_0$ is the radiation length, 14 cm, while $E_c$ is the critical energy for electrons, 32 MeV. Numerically, $l_{max} =48$ cm for $E=1$ GeV and 68 cm for $E=3$~GeV.  Note that at a few GeV, there are large event-to-event fluctuations in shower profiles and the distances to shower maximum (see Fig.~6 in Ref.~\cite{Li:2015kpa}). Importantly, the linear extent of the whole shower is typically less than two meters. In fact, for 1 GeV electrons on average 97\% of energy is deposited within a distance of 1.5 m from the primary vertex. For 3 GeV the corresponding distance is 1.7 m. Thus, the vast majority of the $\nu_{e} + ^{40}$Ar CC events are contained inside the DUNE far detector.  

If all charges deposited in the shower region could be collected, one would measure the starting electron energy by simple calorimetric energy reconstruction.  Complications arise towards the end of the shower, however, where numerous very short tracks are produced by soft electrons away from the main shower region (propagated out by gammas).  Typically, this results in a ``spray" of small ionization-charge deposits around the main tracks as, for example, can be seen in Fig.~\ref{fig:neutrino_event}, where orange dots surround an electromagnetic shower originated by the charged pion. Similar deposits are observed at the end of the muon track, where it decays to a Michel electron and two neutrinos (invisible).
Lastly, they are also ubiquitously created in neutron propagation, as will be discussed below. To fix the terminology we will henceforth refer to this common topology as the ``spray." 

How much of the energy in the spray can be recovered is presently an open question \cite{Reichenbacher2017talk,Asaadi2017talk}. This task places heavy demands on the detector performance and the quality of reconstruction. However, it is certainly not beyond the realm of possibility. In fact, an important proof-of-principle experimental observation of this phenomenon \emph{already exists}, thanks to the work by the ArgoNeuT collaboration~\cite{Lepetic2018APStalk}.  Moreover, very recently, an ArgoNeuT paper~\cite{Acciarri:2018myr} demonstrated that even hits from 0.5 MeV recoil electrons can be detected with 50\% efficiency. Moreover, at 0.8 MeV the energy resolution is as good as 14\%. Therefore, we will include in our analysis the possibility that small ionization deposits (hits) above a certain threshold value can be collected.

Conversely, even with hit-finding thresholds as low as 100 keV some of the spray energy escapes detection. In fact, our simulations show that, in a typical electron shower of 2.4 GeV, as much as 150 MeV of energy goes to spray electrons with energies $<$10 keV. It is, therefore, clear that the spray represents one of the major channels of non-hermeticity for liquid argon detectors.

Lastly, we notice that, when the spray charge deposits are created by nonrelativistic electrons, they are further subjected to large charge recombination corrections. This further decreases the recoverable charges. We discuss the recombination phenomenon below, in the context of proton propagation.

\textit{Gamma rays}. Gamma rays can be created in a variety of processes. As already described, they are copiously produced in electromagnetic showers, by electron bremsstrahlung. If their energy is above 10~MeV~\cite{XCOM}, $\gamma$ rays mostly pair produce, with the radiation length being 14~cm~\cite{PDG}, thus contributing to further development of the shower. At lower energies, $\gamma$ rays undergo energy loss via Compton scattering on electrons in the medium. The signature in this case is a ``spray" of isolated, low-energy ionization deposits, as already mentioned.

Another production mechanism of $\gamma$ rays is via nuclear de-excitations. The resulting $\gamma$ rays are in the MeV energy range and therefore, again, manifest with the Compton ``spray" signature. It is the process of nuclear de-excitations that gave rise to the spray signature observed by the ArgoNeut collaboration~\cite{Lepetic2018APStalk}. 

Lastly, gamma rays are also produced in the decays of neutral pions, $\pi^0\rightarrow\gamma\gamma$. These $\gamma$'s pair convert and, therefore, seed electromagnetic showers. The situation is similar to the electron case considered above, except that the initial $\gamma$'s typically have energies in the 10--100 MeV range, as opposed to a GeV electron. 

\begin{figure*}[t]
    \centering
    \includegraphics[width=0.65\columnwidth]{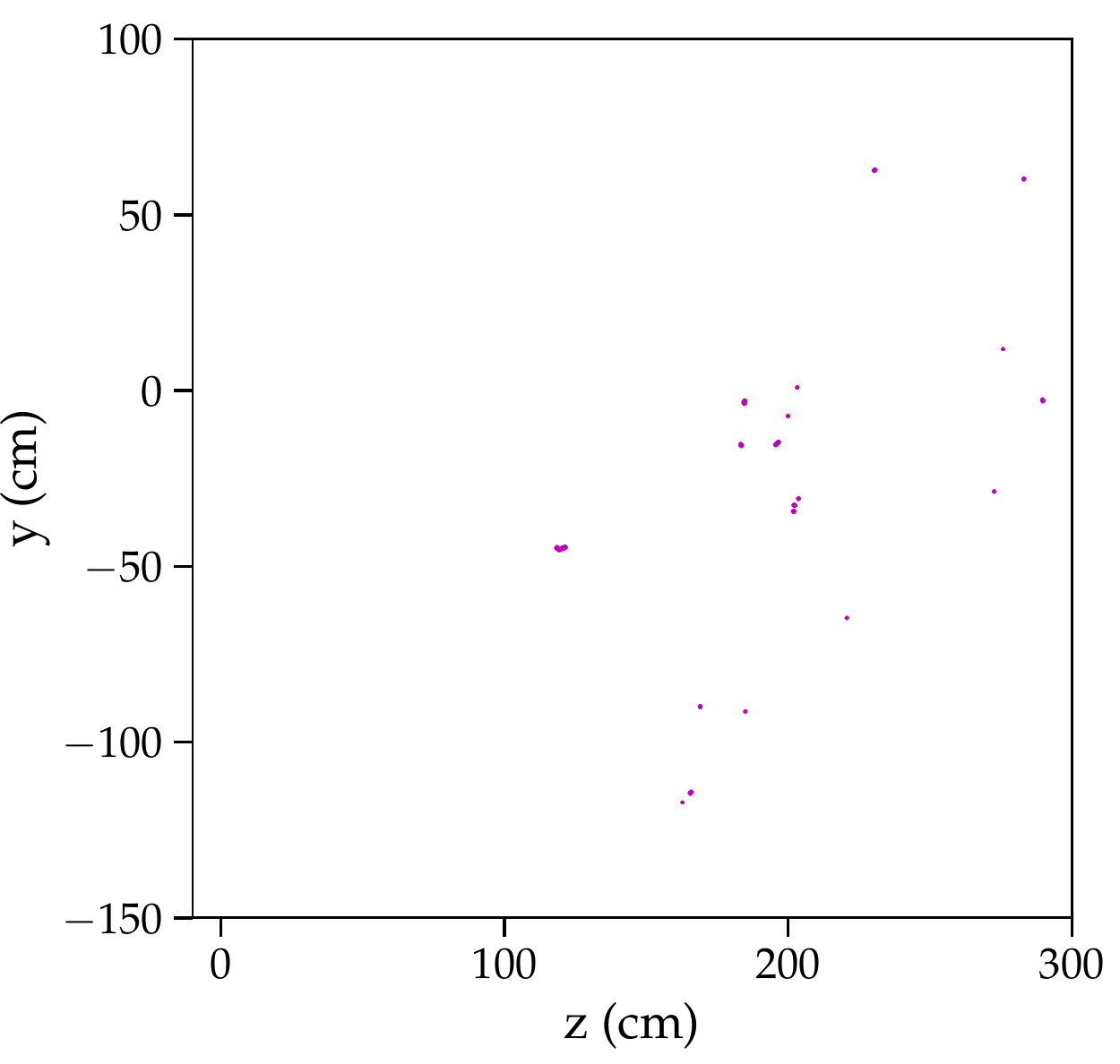}
    \includegraphics[width=0.65\columnwidth]{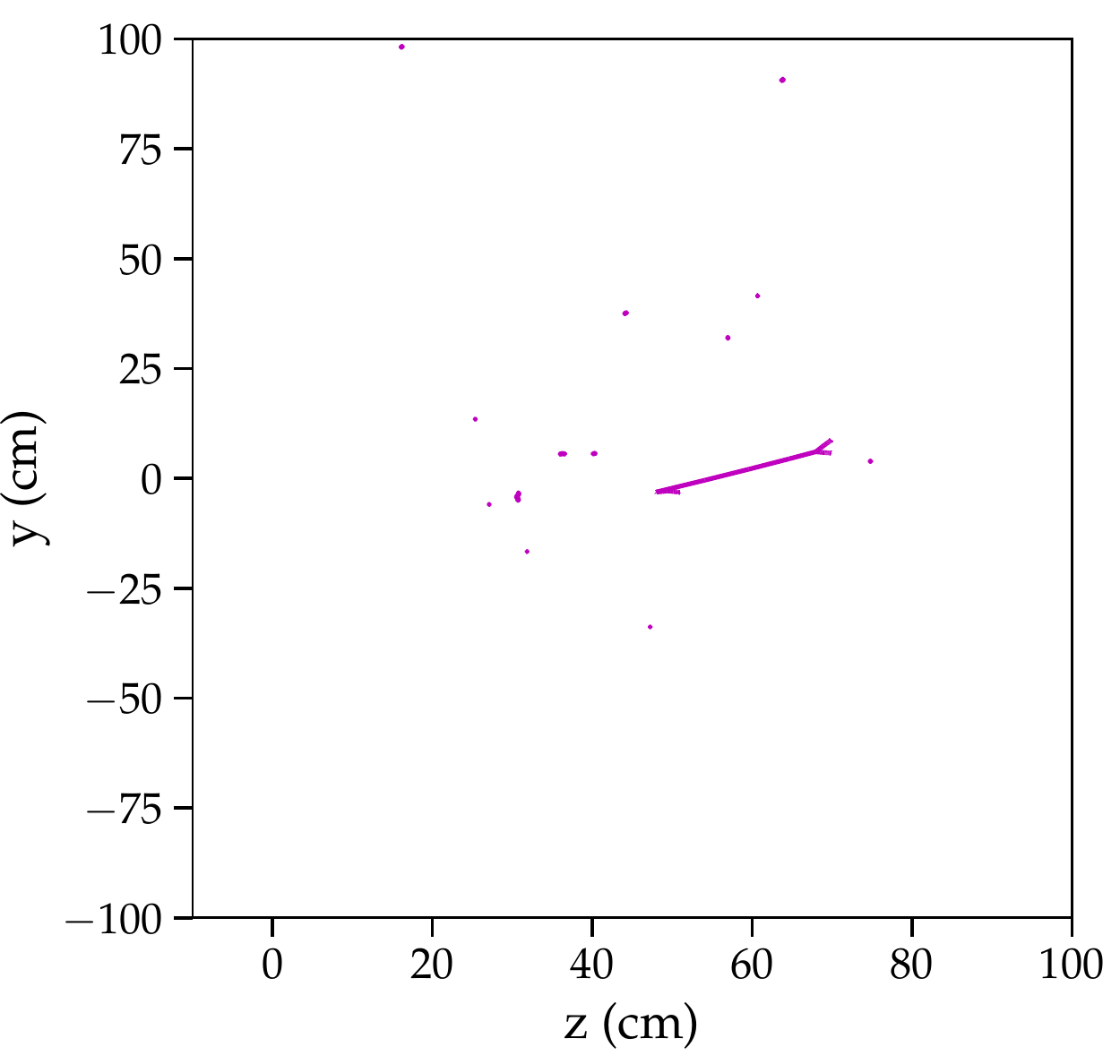}
    \includegraphics[width=0.65\columnwidth]{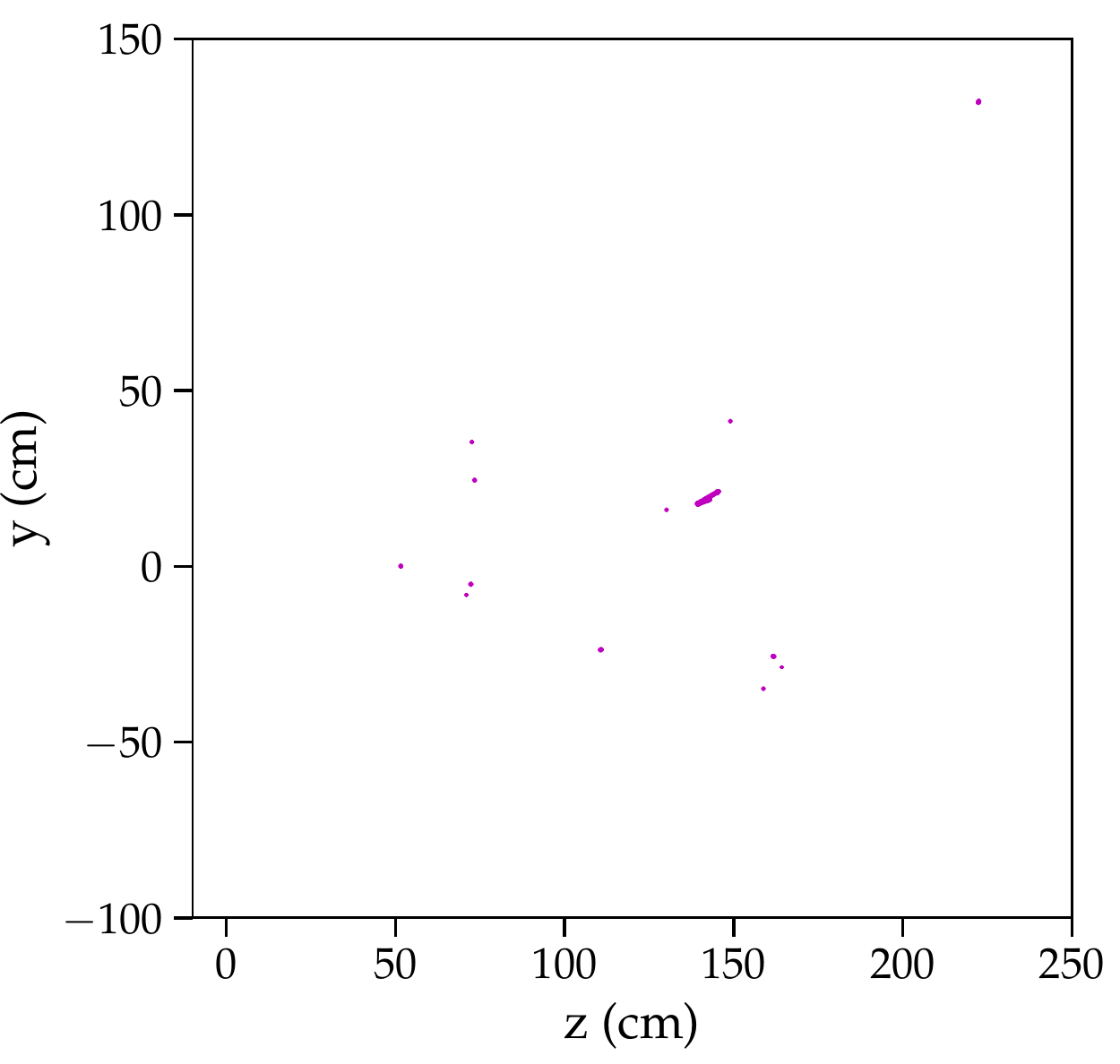}
    \caption{Examples of 500 MeV neutron events.  The neutrons are injected at (0, 0).}
    \label{fig:neutron_events}
\end{figure*}


\subsubsection{Charged hadrons}

Let us now discuss protons and charged pions. Both lose energy continuously through ionization and, just like muons, leave tracks in liquid argon.  There are, however, two important differences between these particles and muons. 

The first one is specific to protons. Because they are often nonrelativistic at DUNE energies, due to their large mass, they have a higher ionization rate than muons or pions. This follows from the Bethe equation, $\langle -dE/dx \rangle \simeq \text{const} \times \beta^{-2}[\ln(2 m_e c^2/I)+2\ln(\beta\gamma)-\beta^2]$, where $I$ is the mean excitation energy (e.g., Ref.~\cite{Jackson:1998nia}). The leading $1/\beta^2$ dependence in the nonrelativistic limit can be simply understood: (i) momentum imparted to a given electron in the medium is proportional to the time the ionizing particle spends in its vicinity, which in turn is inversely proportional to the particle's velocity; (ii) the final kinetic energy transfer to the electron is proportional to the square of the gained momentum.

Numerically, one finds $-dE/dx\sim 12$~MeV/cm for 50~MeV protons (which is the published DUNE threshold; see Table~\ref{tab:threshold}). This can be compared to the loss rate by relativistic muons and pions, $-dE/dx\sim2$ MeV/cm. Because of this, proton tracks tend to be comparatively short. 

The higher density of ionization charge leads to a higher charge recombination rate. The rates for this process were measured by the ICARUS collaboration \cite{Amoruso:2004dy} and we follow their treatment here (see also Ref.~\cite{Sorel:2014rka}). Without recombination, ionization energy loss $\Delta E$ can be converted to charge according to $Q=\Delta E/W_l$, where $W_l = 23.6^{+0.5}_{-0.3}$ eV is related to the energy required to ionize argon. In reality, a fraction of the produced charge, $r$, will recombine, so that $Q=(1-r)\Delta E/W_l$. This fraction depends on the drift field, $\cal E$, and on the charge density, i.e., on $dE/dx$. Following Refs.~\cite{Amoruso:2004dy,Sorel:2014rka}, we apply Birks' law, according to which 
\begin{equation}
    1-r=\frac{0.8}{1+k_Q dE/dx}.
\end{equation}
Here the constant $k_Q= 0.0972$ g/MeV cm$^2$, as measured by ICARUS in a ${\cal E}=500$ V/cm drift field.

Lost charge gives rise to recombination luminescence. While some of it will be quenched (lost to heat due to collisions of argon with impurities, e.g., nitrogen), a fraction will give rise to visible scintillation  (for details see Ref.~\cite{Chiu:2013pwa}). If this light could be accurately measured by the photodetection system, the amount of recombination could be inferred, resulting in improved energy reconstruction, as discussed in Ref.~\cite{Sorel:2014rka}. 

The second difference is the possibility of hadronic (inelastic) scattering, which affects both pions and protons. An energetic hadron can knock several softer particles out of the nucleus, some of which could be neutrons (see later). The typical energy exchange in this process is large. The relevant cross section is almost particle-independent and energy-independent---about the size of the argon nucleus. The corresponding interaction length in liquid argon (density 1.396 g/cm$^3$) is $\sim$1~m, as can be easily estimated. This is shorter than the distance over which a charged pion of energy $\gtrsim$200 MeV would come to rest by ionization losses only. Therefore, hadronic scattering is essential to the transport and energy loss of charged pions at DUNE. It also affects protons, with the relevant crossover energy being higher, $\sim$300 MeV. 

As an illustration, consider, once again, Fig.~\ref{fig:neutrino_event}. The upper track is initially a 780~MeV proton, which undergoes an inelastic interaction around $z \simeq 40$ cm. The final state contains three protons, with kinetic energies of 250, 200 and 150 MeV. The rest of the initial proton energy goes into ionization prior to the collision and to nuclear breakup at the hadronic interaction point. The track just below is a 1.7 GeV $\pi^+$, which inelastically interacts at $z \simeq 143$ cm. This interaction creates a 60 MeV $\pi^0$,  a 500 MeV $\pi^+$, a 300 MeV deuteron, a 100 MeV proton, four neutrons, three alpha particles, and a remnant neon nucleus. It is closely followed by another hadronic interaction of the 
$\pi^+$, at $z \simeq 148$ cm. This interaction creates a number of hadrons, including several energetic neutrons that will be discussed in the next subsection.

Once charged pions come to rest, their fate depends on their charge.  Negatively charged pions would bind with an argon nucleus, capture on it, and break it up.  This way, the pion rest mass turns into the kinetic energy of the outgoing nucleons and nuclear binding energy. In contrast, the $\pi^+$ decays, $\pi^+\rightarrow\mu^+ +\bar\nu_\mu$, followed by the decay $\mu^+\rightarrow e^+ +\bar\nu_e+\nu_\mu$. The neutrinos are not observable, and are another source of missing energy.


\subsubsection{Neutrons}

As noted before, neutrons deserve a separate treatment. Being electrically neutral, they themselves do not leave ionization tracks. Above tens of MeV, neutrons interact hadronically, with a cross section comparable to that of protons.  In this regard,  neutrons are similar to protons and charged pions.  Below tens of MeV, neutrons lose energy either by producing more soft neutrons, or by scattering off nuclei and leaving them in excited states~\cite{Brown:2018jhj}.  In the process, de-excitation energy is deposited in the charge ``spray" form (by Compton scattering of gamma rays on numerous electrons).  Ultimately, low-energy neutrons can travel as far as several meters from the injection point vertex.

Only a fraction of the initial neutron energy can in principle be recovered through ionization: the part that goes into secondary and tertiary charged particles and gamma rays created in neutron collisions with argon nuclei. The efficiency with which this ionization charge can be recovered depends on the details of the event and the performance of the detector. Understanding this physics is critical for accurate modeling of the energy resolution at DUNE. 

As the first illustration, let us again turn to the event in Fig.~\ref{fig:neutrino_event}. 
The event contains several neutrons. Two of them originate at the primary vertex and have initial kinetic energies of 120 and 50 MeV. In addition, the proton collision at $z \simeq 40$ cm knocks out six neutrons with a total energy of 20 MeV, while the charged pion collision at $z \simeq 143$ cm knocks out four neutrons, with energies  48, 5, 4, and 3 MeV. The charge pion then has another hadronic interaction, at $z \simeq 148$ cm, which knocks out an 81 MeV neutron, a 30 MeV neutron, together with some MeV soft neutrons. We see that neutrons can originate both at the neutrino interaction vertex and in secondary interactions and can carry a non-negligible fraction of the energy in an event. For the case at hand, we have 170 MeV in primary neutrons and 230 MeV in secondary neutrons.

The magenta points mark charge deposits that are traced back to neutrons. Notice that there are no ionization tracks. The distribution has the characteristic ``spray'' topology one expects from Compton scattering of low-energy de-excitation gamma rays. It is noteworthy, however, that in this case the ``spray'' region has a linear extent of over a meter. Since this is considerably more than the gamma-ray radiation length, the deposits must be created by gamma rays from a number of distributed sources. Indeed, neutrons break up a number of argon nuclei as they random walk over distance scales of meters from their creation points. This nuclear breakup process is a major channel of neutron energy loss; only a fraction of the deposited energy is eventually converted into ionization (as the break up products de-excite into the nuclear ground state). Indeed, by adding up the charge in the magenta points, one recovers only 250 MeV out of the total 400 MeV of the neutron energy.

As the next illustration, consider Fig.~\ref{fig:neutron_events}, which shows three examples of 500 MeV neutron events from the {\tt FLUKA} simulation. In all three cases, the initial state is the same: the neutron is injected at point (0, 0), with the initial momentum pointing along the $z$ axis. One can immediately see that charge deposits due to neutrons have complicated and fluctuating topology. In particular, in addition to the ``spray" of many small charge deposits, neutrons can knock out protons energetic enough to leave distinct tracks, as most clearly seen in the middle panel. These protons can be far from the neutron injection point (0.5--1.5 m or more) and may not point accurately back to it. They are also nonrelativistic and as such subject to a large charge recombination correction.

\begin{figure}[t]
    \centering
    \includegraphics[width=\columnwidth]{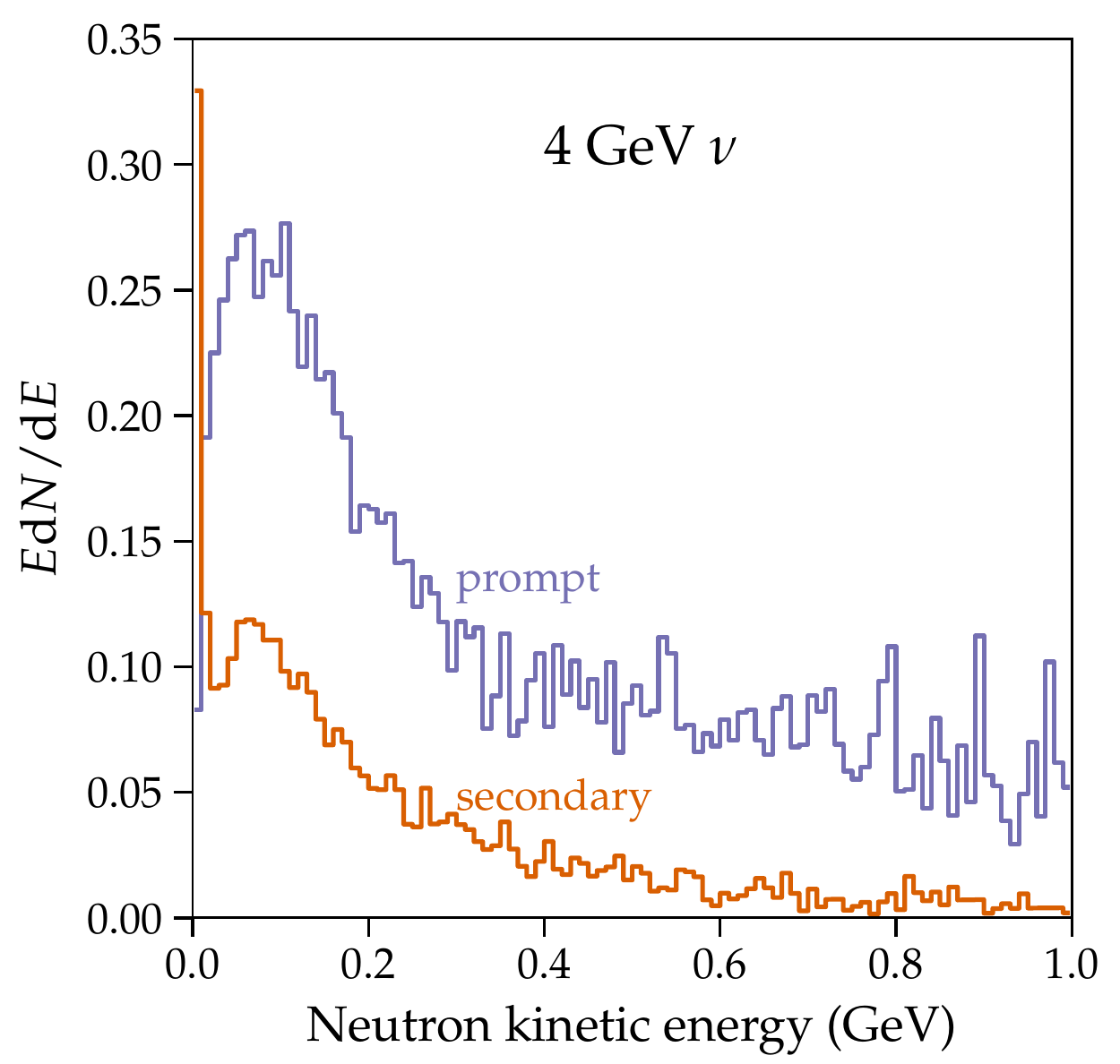}
    \caption{Energy distribution of the neutrons in 4 GeV neutrino interactions. The top histogram corresponds to prompt neutrons, which come from the nucleus struck by the neutrino; the bottom one corresponds to secondary neutrons, which are knocked out of other argon nuclei as the event develops in the detector.}
    \label{fig:neutron_distribution}
\end{figure}

\begin{figure}[t]
    \centering
    \includegraphics[width=\columnwidth]{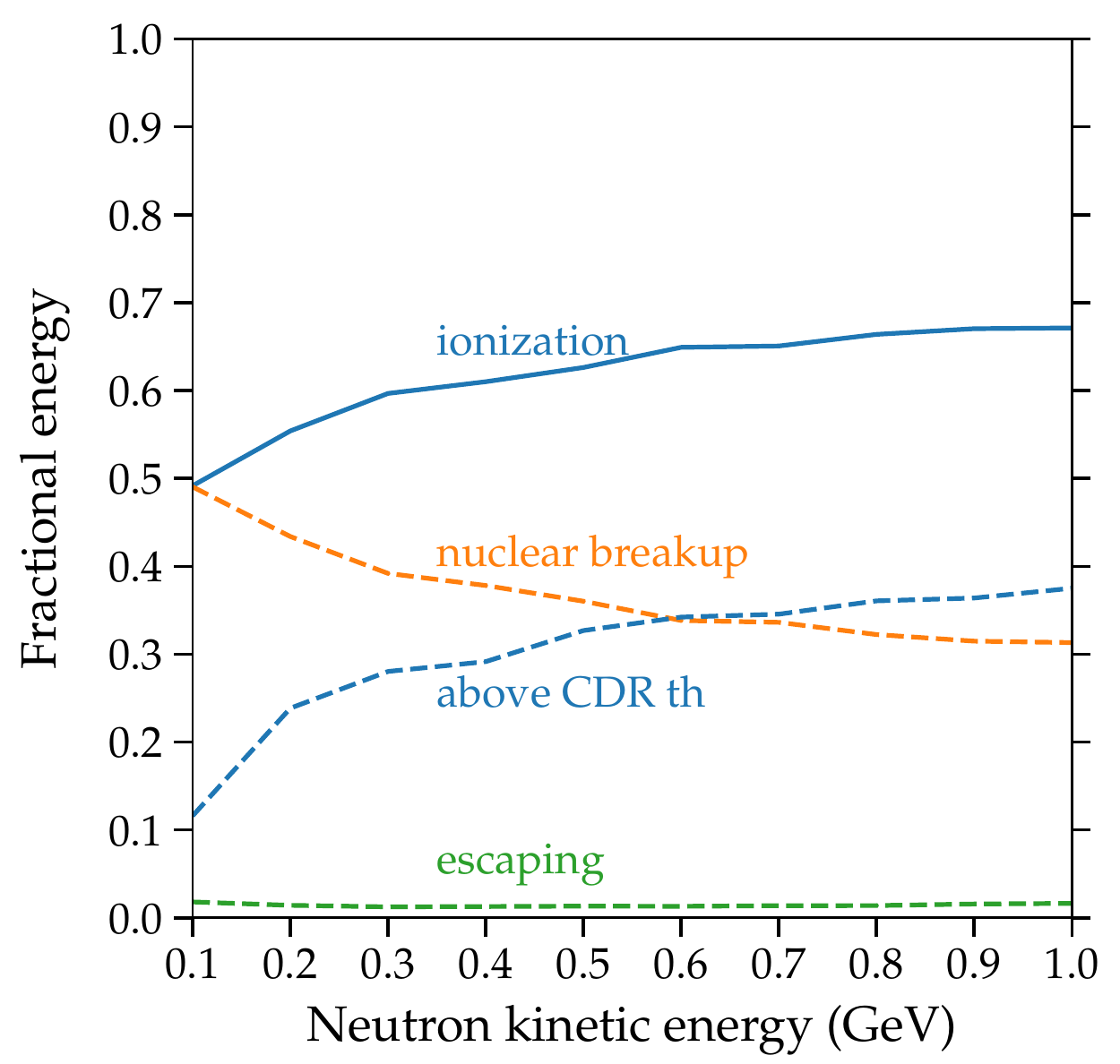}
    \caption{Energy breakdown of neutrons at different energies. Shown are the energy fractions that go to ionization (\emph{solid blue}) and to nuclear breakup (\emph{dashed orange}). Also shown is the fraction of the ionization energy in particles above the CDR thresholds (\emph{dashed blue}). The ionization charges are given before charge recombination.}
    \label{fig:neutron_budget}
\end{figure}

Because of this complexity and large variability, detailed modeling of neutrons is essential for understanding the performance of DUNE. This is one of the goals of this paper. 
Figure~\ref{fig:neutron_distribution} shows the energy distributions in both prompt and secondary neutrons in 4 GeV neutrino interactions. The average energy of a prompt neutron is 160 MeV, although the distribution is broad and values as high as 1 GeV contribute.
Figure~\ref{fig:neutron_budget} shows the breakdown of the average energy deposition by neutrons into ionization vs. nuclei breakup, as a function of neutron kinetic energy.  We see that at 200--300 MeV, the invisible nuclear breakup takes up  $\gtrsim$40\% of the neutron energy.  This fraction is large because neutrons, by inelastically interacting with argon nuclei, tend to produce more neutrons, and this cascade ends up disrupting a large number of nuclei in the medium.

Collecting the $\sim$60\% of potentially visible energy is not straightforward.  Indeed, it is distributed in a combination of ``spray" and isolated proton tracks. As an illustration, if we impose the DUNE CDR thresholds on the events (Table~\ref{tab:threshold}), we find that the solid blue curve reduces to the dashed blue one in Fig.~\ref{fig:neutron_budget}, i.e. that half of all charge will be missed.

We close this section with comments on containment and timing. While the majority of scattering events in the geometry of the DUNE far detector should be contained, some fraction will have one or more escaping particles. This especially applies to muons produced in the scattering of $\nu_\mu$, a third of which could be escaping, as already mentioned. Because the energy resolution of such escaping muons is much poorer, adding them to the general sample will significantly degrade the average resolution. This, however, introduces an artificial complication; after all, such events are easily distinguished from the contained ones in the experiment. If a study requires good energy resolution, the exiting events can be dropped, with a modest loss of statistics. In general, for maximal flexibility, analyses should keep both, but divide the data set into contained and exiting, never averaging the two.  Because our primary focus in this paper is on understanding the physics of the energy loss of the liquid argon technology, we will henceforth focus on contained events. Particle leakage should be kept in mind, however, when considering smaller detectors, including the DUNE near detector.

As for the timing of charge deposition, one may wonder if neutron-induced charges are delayed compared to the prompt event.  While the entire neutron-induced cascade can take a long time to develop, we have explicitly checked that, for neutrons with initial energies in the hundreds of MeV, the peak of energy loss occurs $10^{-8}$~s after creation. This can be understood simply as the time scale of hadronic interactions. By $10^{-6}$~s, most of the original neutron energy has been dissipated. Therefore, one can use timing cuts to associate neutron energy deposits with the main event.

\begin{figure*}[tbh]
	\begin{center}
        \includegraphics[width=\columnwidth]{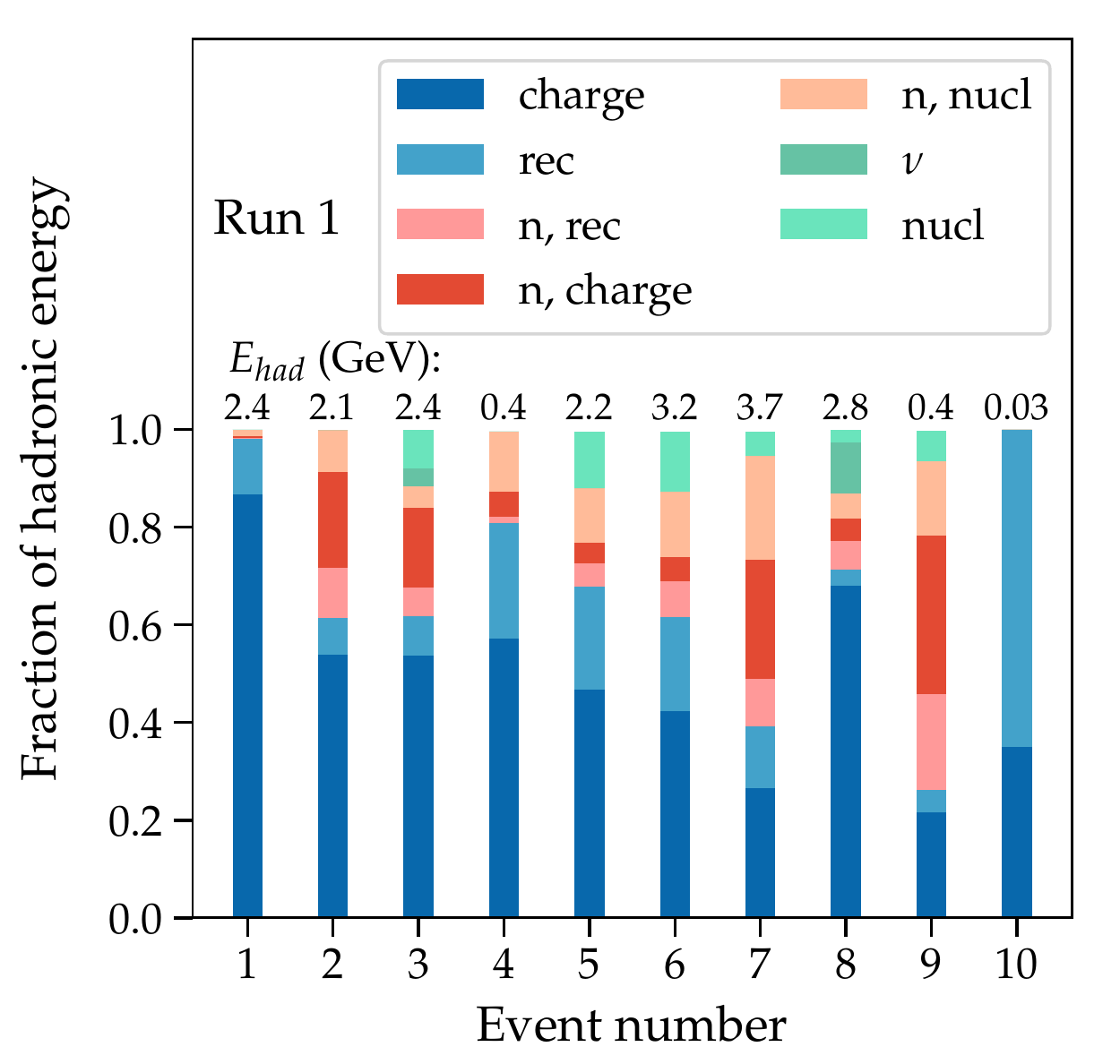}
        \includegraphics[width=\columnwidth]{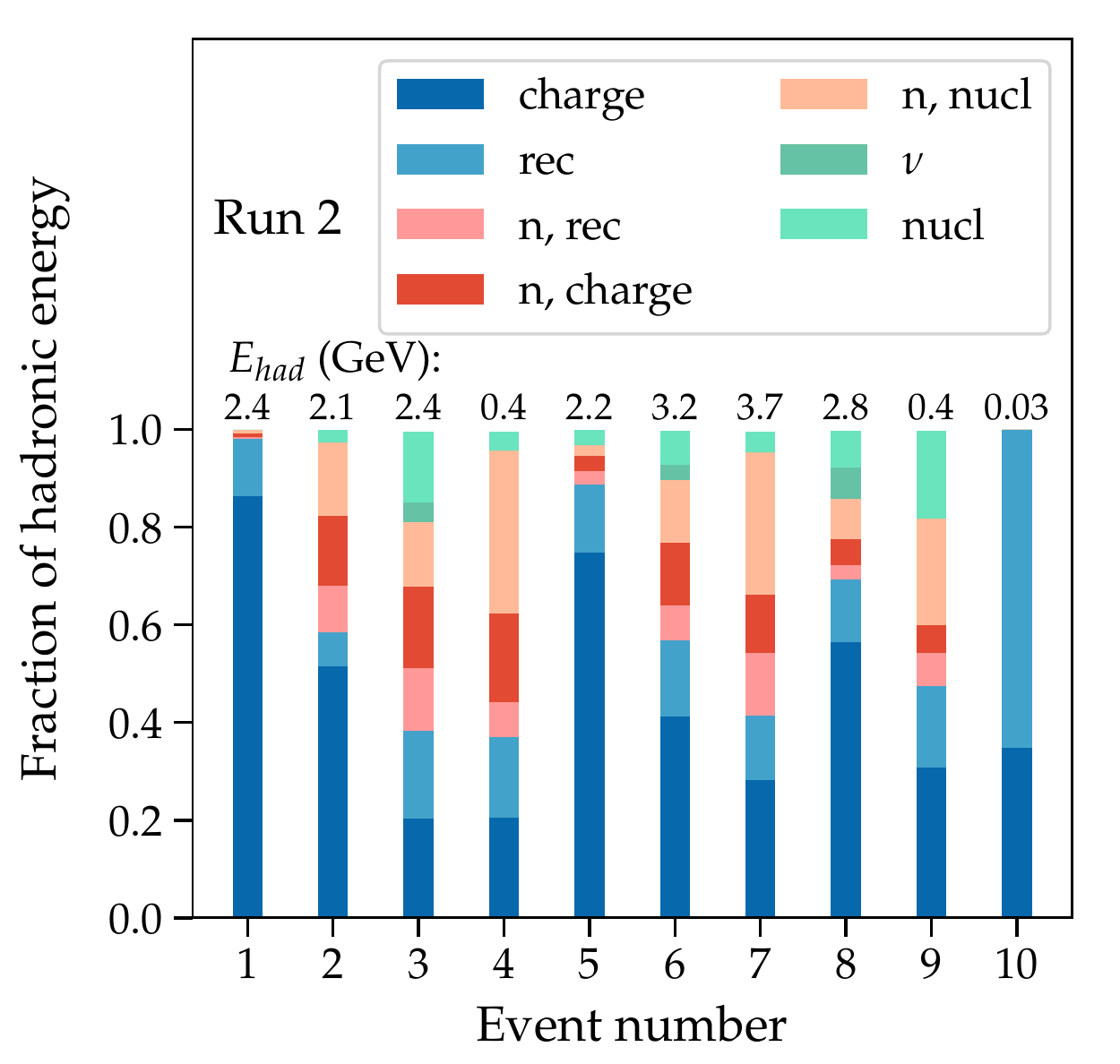}
        \caption{The final energy budget of the first ten events from {\tt GENIE}, depicted earlier in Fig.~\ref{fig:hadronic_fluctuation}. The left and right panels show two {\tt FLUKA} realizations. Here, ``charge" reflects the total ionization produced, ``rec" refers to charge lost to recombination, ``nucl" is the energy that goes into the breakup of argon nuclei in the detector and ``$\nu$" is the energy carried away in neutrinos from $\pi/\mu$ decays. The corresponding processes involving neutrons are labeled separately. }
        \label{fig:two_FLUKA_realizations}
    \end{center}
\end{figure*}


\section{Simulation results}
\label{sect:results}

In the preceding section, we have identified a number of physical processes that can affect the hermeticity of the DUNE detector. Let us now summarize them. In a liquid argon detector, energy can be missed in the following channels:
\begin{enumerate}
    \item at the end of electromagnetic showers---in the spray of small charge deposits from Compton electrons;
    \item via increased charge recombination that occurs on tracks of nonrelativistic particles, primarily protons;
    \item in the process of nuclear breakup;
    \item via neutrinos from meson and muon decays;
    \item as subthreshold particles produced at the various stages of the event development;
    \item via primary and secondary neutrons.
\end{enumerate}
Strictly speaking, the last channel can be viewed as a subset of the earlier ones: neutron energy can be missed because of nuclear breakup, in the Compton spray, or via the recombination of the ionization charge from the secondary charged particles. Yet, from the practical point of view, in all these cases, neutrons present a special challenge. The energy fraction that goes to nuclear breakup is much higher for neutrons than for charged particles. The spray region from neutrons can be meters in extent and could be difficult to reconstruct. Finally, protons created by neutrons are ``detached" from the main event and may not be efficiently picked up by reconstruction algorithms. It therefore makes sense to classify neutrons in a standalone category. 

Our next tasks are to quantify, using our simulation framework: 
\begin{itemize}
    \item how much energy, on average, is lost in each channel,
    \item how much variability is exhibited by each channel, and
    \item how these factors combine to yield the energy resolution of the detector.
\end{itemize}

The results will clearly depend on 
\begin{itemize}
    \item the assumed detection thresholds for each particle species, 
    \item whether recombination effects can be corrected for (i.e., the quality of event reconstruction), and 
    \item whether we are dealing with neutrino- or antineutrino-induced events.
\end{itemize}

We begin by illustrating the relevant physics with a small set of events. We then quantify the average energy losses and discuss the fluctuations in the relevant channels.


\subsection{Example events}
\label{sect:example10}

Let us return to the discussion at the end of Sec.~\ref{sect:primary_interactions}. There, we described the challenge of trying to predict the amount of missing energy in an event by considering only the primary interaction. We can now restate the problem in the context of our present discussion, by noting that prompt particles do not directly correspond to the missing energy channels we just enumerated. To make the required connection, one needs to simulate the full propagation of all particles through liquid argon. We turn to this next.

We again find it convenient to start with an illustration. We recall the 10 events depicted in Fig.~\ref{fig:hadronic_fluctuation}. Upon running them through our full simulation pipeline, we obtain the results shown in Fig.~\ref{fig:two_FLUKA_realizations}. Notice that the two panels show two different {\tt FLUKA} simulations of \emph{the same 10 initial states}. As before, the total hadronic energies are labeled above each bar. The color partitioning of the bars reflects the fractions of the energy going to ionization charge (\emph{blue}) and to different missing energy channels (\emph{other colors as labeled in the legend}) in each case. The processes involving neutrons are labeled separately, including the total ionization charge created by them (\emph{red}). Detection thresholds have not yet been imposed at this point.

Examining Figs.~\ref{fig:hadronic_fluctuation} and \ref{fig:two_FLUKA_realizations} together, one immediately notices several important features.
\begin{itemize}
    \item The fraction of hadronic energy that goes to ionization charge varies greatly from event to event. Even for events 1 and 3, which have the same $E_\text{had}=2.4$~GeV, the missing energy is different. Physically, this is because the first one is dominated by electromagnetic showers, while the second one involves propagating charged hadrons ({\it cf}. Fig.~\ref{fig:hadronic_fluctuation}).
    \item There is no clearly dominant missing energy channel. Other than the losses from decay neutrinos, which are always small, all other channels can be significant. Which one dominates depends on the composition of the prompt system and, sometimes, on the stochastic factors in propagation.
    \item Neutrons can carry a sizable fraction of the event's energy. As an example consider event 7, where the hadronic system has 3.7 GeV, almost all of the original neutrino's energy. In the left panel, less than a third of that energy goes into ionization not created by neutrons.
    \item Variations between different realizations of the same primary events can be large. Compare, e.g., the left and right realizations of event 3, which is dominated by prompt charged pions. Almost all loss channels are larger on the right, as the pions transferred a larger fraction of their energy to neutrons in that run. Notable variability is also observed in other events with the energetic hadronic system, such as 5 or 7.
\end{itemize}


\subsection{Average losses}
\label{sect:averagelosses}

\begin{figure}[t]
	\begin{center}
        \includegraphics[width=\columnwidth]{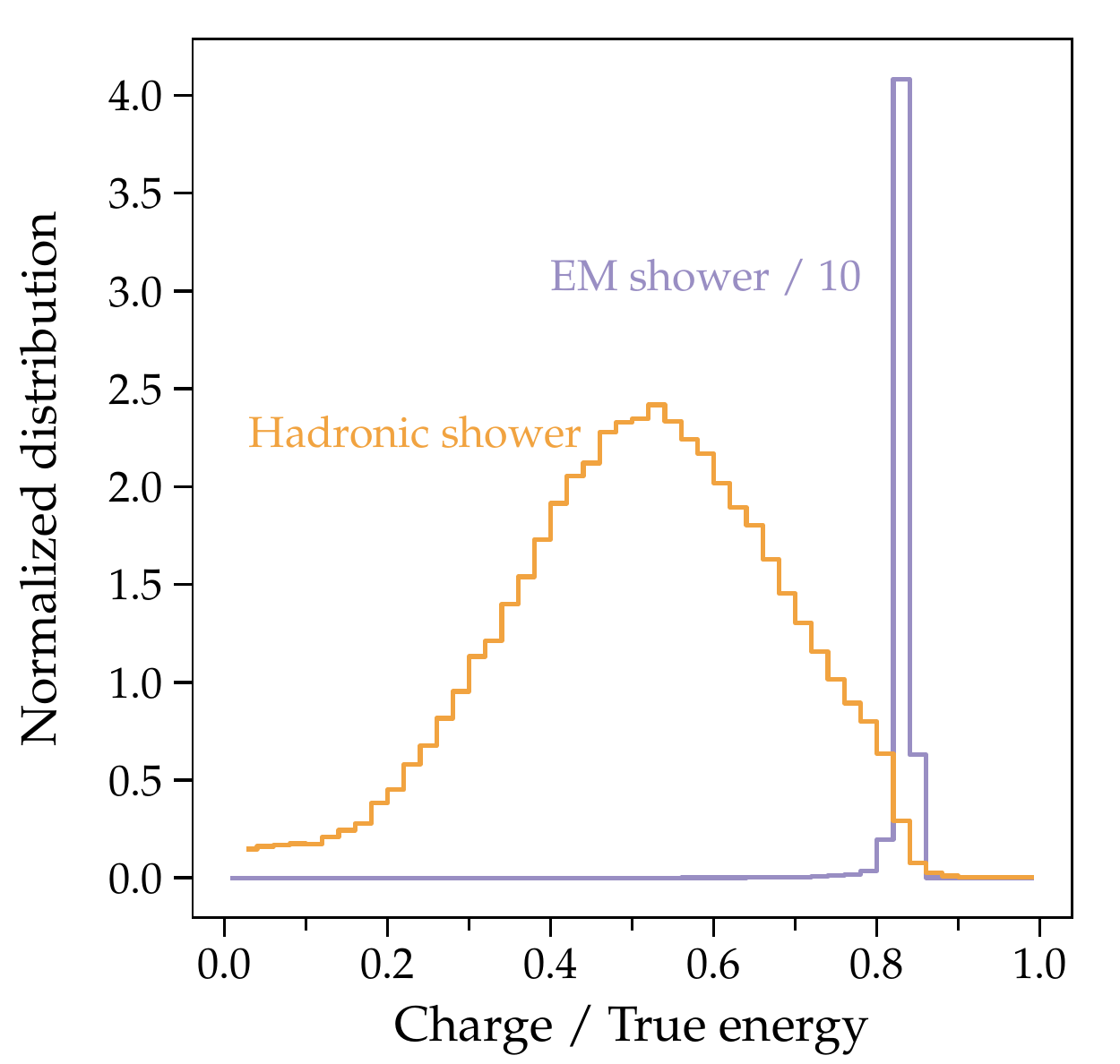}
        \caption{Histograms of the energy fraction that goes to ionization charge, for hadron- and electron-induced showers. Initial neutrinos have a flat spectrum in the $[2,4]$ GeV range.}
        \label{fig:charge_distributions}
    \end{center}
\end{figure}

\begin{figure}[t]
	\begin{center}
        \includegraphics[width=\columnwidth]{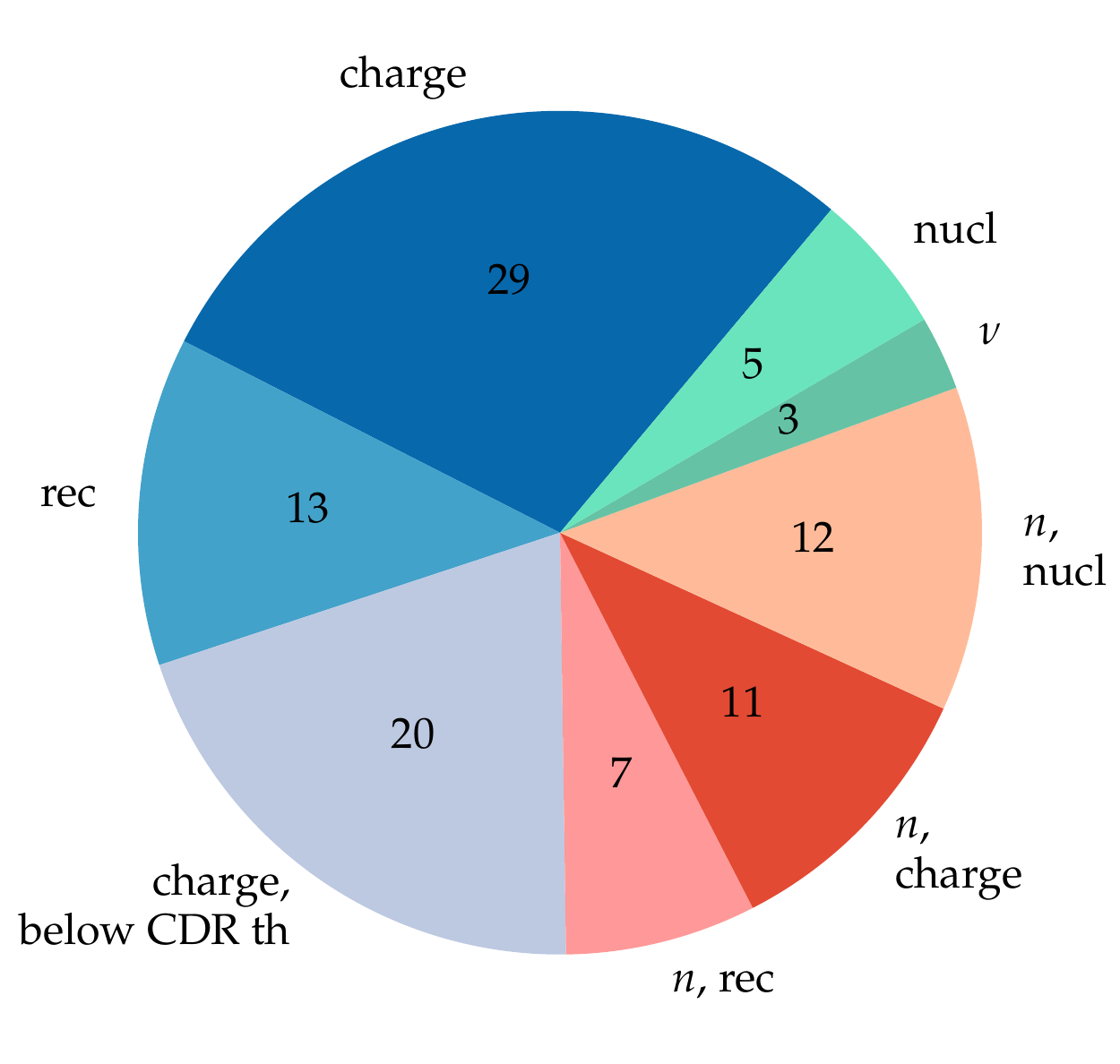}
        \caption{Hadronic energy budget after fully propagating neutrino events ({\it cf}. Fig.~\ref{fig:hadronic_budget_primary}). The averaging was performed over a set of 10,000 4 GeV $\nu_\mu$ CC interactions. Shown are the fractions of the hadronic energy that go into ionization charge (\emph{charge}) above and below the CDR thresholds, that are lost to recombination (\emph{rec}), lost to nuclear breakup (\emph{nucl}) and, finally, that escape as decay neutrinos (\emph{$\nu$}). As in Fig.~\ref{fig:two_FLUKA_realizations}, the corresponding processes for neutrons are shown separately (\emph{n}).}
        \label{fig:missing_energy_4GeV}
    \end{center}
\end{figure}

\begin{figure*}[t]
	\begin{center}
        \includegraphics[width=\columnwidth]{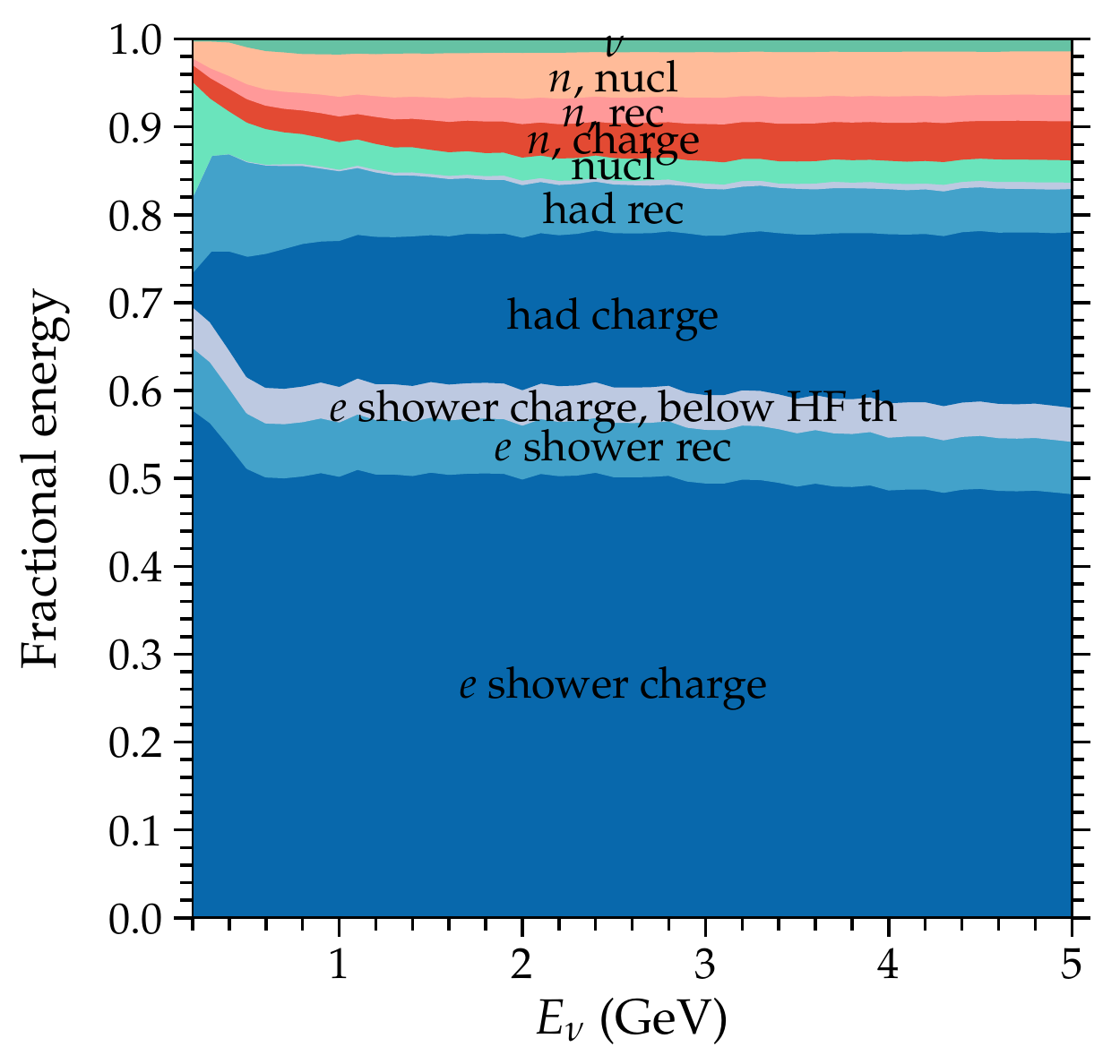}
        \includegraphics[width=\columnwidth]{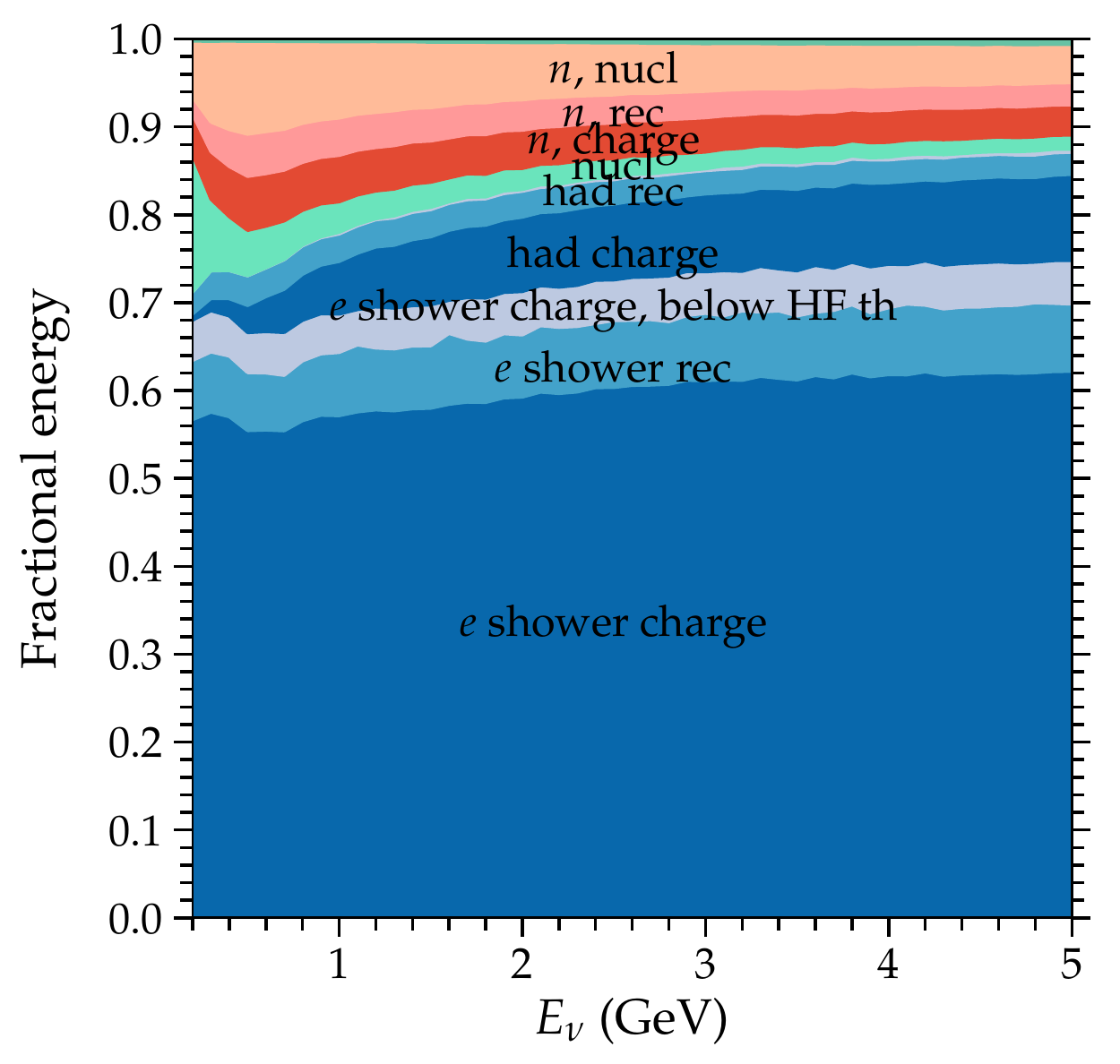}
        \caption{The energy budget for $\nu_e+ ^{40}$Ar (\emph{left panel}) and $\bar\nu_e+ ^{40}$Ar (\emph{right panel}) scattering events as a function of the (anti)neutrino energy. Both the electromagnetic shower from the final-state electron and the hadronic system are included. The hadronic energy channels are the same is in Fig.~\ref{fig:missing_energy_4GeV}. In contrast to Fig.~\ref{fig:missing_energy_4GeV}, however, hit-finding thresholds have been applied here (see text).}
        \label{fig:missing_energy}
    \end{center}
\end{figure*}

\begin{table}[b]
\caption{Outcomes of neutron propagation in liquid argon detectors. The rows correspond to different starting neutron kinetic energies, $T_{n}$. The quantity $E_{\rm vis}$ is the total average energy that goes to ionizing particles in each case. Individual contributions are given for  protons ($E_{p}$), electrons ($E_{e}$), and ions ($E_{\rm ion}$).  The corresponding columns with ``$Q$'' incorporate recombination corrections (and thus are in proportion to the actual ionization charges). All numbers are given in GeV.}
\begin{center}
\begin{tabular*}{0.47\textwidth}{@{\extracolsep{\fill} } c|cc|cccccc}
\hline
 $T_n$ &  $E_{\rm vis}$ &  $Q_{\rm vis}$ &   $E_{p}$ &  $Q_{p}$ &      $E_{e}$  &     $Q_{e}$ &    $E_{\rm ion}$   &   $Q_{\rm ion}$ \\ \hline 
0.1  &  0.049  &  0.025  &  0.015  &  0.006  &  0.020  &  0.017  &  0.014  &  0.003 \\
0.2  &  0.111  &  0.058  &  0.051  &  0.025  &  0.032  &  0.026  &  0.028  &  0.006 \\
0.3  &  0.179  &  0.099  &  0.096  &  0.054  &  0.044  &  0.036  &  0.039  &  0.009 \\
0.4  &  0.244  &  0.138  &  0.129  &  0.075  &  0.061  &  0.050  &  0.051  &  0.012 \\
0.5  &  0.313  &  0.182  &  0.164  &  0.098  &  0.076  &  0.063  &  0.066  &  0.016 \\
0.6  &  0.389  &  0.235  &  0.196  &  0.120  &  0.097  &  0.080  &  0.084  &  0.024 \\
0.7  &  0.455  &  0.277  &  0.229  &  0.141  &  0.115  &  0.096  &  0.094  &  0.025 \\
0.8  &  0.531  &  0.331  &  0.257  &  0.161  &  0.145  &  0.121  &  0.107  &  0.030 \\
0.9  &  0.603  &  0.381  &  0.298  &  0.189  &  0.160  &  0.134  &  0.115  &  0.031 \\
1.0  &  0.671  &  0.427  &  0.316  &  0.200  &  0.186  &  0.155  &  0.129  &  0.034 \\
\hline
\end{tabular*}
\end{center}
\label{table_neutrons}
\end{table}%

Given the large diversity of the possible final states and, furthermore, the stochastic variability of shower development, one may suspect that to quantify both the average energy losses and their fluctuations in each channel a high-statistics sample of events is required. That this is indeed so is illustrated in Fig.~\ref{fig:charge_distributions}, where we show the distributions of the visible energy fraction in $\nu_e+ ^{40}$Ar events generated from a flat neutrino spectrum in the range from 2 to 4 GeV. Electron-induced showers and those induced by the hadronic system are separated in the individual histograms. We see that while in electromagnetic showers the invisible energy fraction is nearly always the same, in hadronic showers it is highly variable, consistent with the behavior observed for the ten events of Sec.~\ref{sect:example10}. A lot of statistics is then required to fully characterize it.

Moreover, since the loss fractions may change with the incoming neutrino energy, large samples must be generated for each energy value of interest. The study in question is thus by necessity computer-intensive. We report its findings below, in several steps.

First, let us specialize to a fixed neutrino energy. To this end, we return to the 10,000 CC neutrino-argon 4 GeV scattering events we previously discussed in Sec.~\ref{sect:primary_interactions}. We run all of them through {\tt FLUKA} and perform averaging over the results. To make the comparison with the earlier discussion straightforward, we also impose the CDR thresholds (according to Table~\ref{tab:threshold}). The difference with Sec.~\ref{sect:primary_interactions} is that now we apply them consistently, to the low-energy particles produced \emph{at all stages} in the events.

The outcome of this analysis is summarized in pie-chart form in Fig.~\ref{fig:missing_energy_4GeV}, which should be compared with Fig.~\ref{fig:hadronic_budget_primary}.  
It is immediately apparent that the new pie chart is qualitatively different: the energy budget now has many components, with none clearly dominating the rest, in agreement with what we already saw in Fig.~\ref{fig:two_FLUKA_realizations}. The contrasts are numerous. For example, loss to nuclear breakup comprises a significant part of the overall energy budget in Fig.~\ref{fig:missing_energy_4GeV}, especially the part caused by neutrons. This category is not present at all in Fig.~\ref{fig:hadronic_budget_primary}. Even among the categories that are common between the two pie charts, there are notable differences. The neutron-related slices in Fig.~\ref{fig:missing_energy_4GeV} together add up to 30\% of $E_\text{had}$, significantly more that the corresponding slice in Fig.~\ref{fig:hadronic_budget_primary}. The difference is made up by  secondary neutrons knocked out in propagation. The fraction of energy that goes to subthreshold losses is as large as 20\% in Fig.~\ref{fig:missing_energy_4GeV}, dramatically larger than the corresponding slice in Fig.~\ref{fig:hadronic_budget_primary}. Clearly, considering full propagation qualitatively changes every aspect of the problem.

The total visible energy in Fig.~\ref{fig:hadronic_budget_primary} is seen to be only 40\%, significantly lower than in Fig.~\ref{fig:missing_energy_4GeV}. We see that this number strongly depends on the experimental performance: it can be as low as 29\% if all neutrons are also missed, a maximally pessimistic scenario, or as high as 60\%, if all neutron-created charge is detected and all thresholds are lowered to zero.

We see that a consistent application of the CDR thresholds to entire events in DUNE leads to dramatic energy losses. In this respect, it should be kept in mind that the CDR thresholds are extremely conservative and in no way reflect fundamental technology limitations. Indeed, as already mentioned in Sec.~\ref{sect:propagation}, liquid argon detectors can observe much less energetic particles \cite{Reichenbacher2017talk,Asaadi2017talk}, even down to MeV-scale deposits from de-excitation gamma rays, as recently shown by ArgoNeuT~\cite{Lepetic2018APStalk}. Accordingly, while lowering thresholds all the way to zero may not be realistic, it is of interest to consider what can be achieved under optimistic assumptions. To this end, if we impose hit-finding thresholds \cite{Reichenbacher2017talk,Asaadi2017talk,Lepetic2018APStalk} of 100 keV (applied to actual ionization charge), the below-threshold slice in Fig.~\ref{fig:hadronic_budget_primary} shrinks from 20\% to as little as 2\%. For the rest of this section, we will adopt such optimistic values.

With this setup, we can now turn to our general results. We repeat the same full event simulations as done before, for a set of neutrino energies in the range of 0.1--5 GeV, which encompasses the spectrum of the DUNE beam. The results are presented in the left panel of Fig.~\ref{fig:missing_energy}. This time, we specialize to $\nu_e+^{40}$Ar scattering and include the electron shower in the overall energy budget, so that the total adds up to the incoming neutrino energy. We also impose low particle thresholds of 100 keV per hit, as discussed earlier. 

While some variation with energy is observed, overall the fractions are quite stable. This result is perhaps surprising, given the significant change of the final-state composition as the neutrino energy is varied between 1 and 5 GeV. The main changes with energy are in the fraction of hadronic energy that gets lost to charge recombination and in the fraction that goes to neutrons.   

Because of its high importance, we carried out a dedicated investigation of the energy dependence of the neutron channels. For this, we simulated complete propagation in liquid argon of neutrons with initial kinetic energies from 100 MeV to 1 GeV, averaging, as before, over 10,000 events at each energy value. Table~\ref{table_neutrons} shows the results of these simulations. The column $E_\text{vis}$ gives the amount of energy that goes to ionization before recombination is taken into account; $Q_\text{vis}$ gives the corresponding answer after recombination (e.g., actual ionization charge). The subsequent columns show the detailed breakdown of energies between the different final-state channels: protons, electrons and heavier ions. Once again, the notation $Q$ is used to denote actual ionization charge.

Lastly, the right panel of Fig.~\ref{fig:missing_energy} shows the energy-dependent missing energy budget for antineutrino scattering. In this case, the fraction of energy that goes to the electron-induced shower is larger, in accord with Fig.~\ref{fig:primary_hadronic}. Another obvious difference with the left panel is the increased fraction of the total energy that goes to neutrons at sub-GeV energies. This, again, is in agreement with our earlier discussion: at these low energies, the CC quasi-elastic process dominates, which for antineutrinos produces a neutron in the final state. It is notable, however, that at $E_\nu \sim$ 3--4 GeV the neutron fraction of the total energy is similar for neutrinos and antineutrinos.


\subsection{Fluctuations and energy resolution}
\label{sect:modelresolution}

With the preceding discussion, we have established and quantified average losses in all channels of non-hermeticity of DUNE. This information allows us to build a reconstruction procedure for neutrino energy. Simply put, knowing how much energy is expected to be lost in each channel, one can work backwards and infer the most likely full energy from the fraction that is visible. As a simple example, even if one was to detect only 29\% of the total hadronic energy, as in Fig.~\ref{fig:missing_energy_4GeV}, as long as one somehow knew it was exactly 29\%, one could reconstruct the true energy. 
In this sense, by modeling missing energy one obtains, effectively, a conversion coefficient between observed ionization charges and true energies. 

Energy loss, however, means information loss and there is a price to pay for it. Broadly speaking, there are two relevant considerations. First, the missing energy fractions need to be known precisely. This means one must accurately calibrate the detector response to all relevant secondary particles and, at the same time, validate the neutrino-nucleus cross section model.
As stressed earlier, our missing energy fractions in this paper are specific to the versions of {\tt GENIE} and {\tt FLUKA} we employ. Model deficiencies in these codes will translate into systematic errors for the energy scale.  

Second, even with perfect modeling of detection physics, one is able to infer the true value of the neutrino energy only \emph{on average}, not on an event-by-event basis. 
Indeed, as we saw in the beginning of Sec.~\ref{sect:example10}, events with the same true hadronic energies can leave very different amounts of visible ionization charge. Even different detector realizations of the same primary interaction can have quite different visible energies (as we learned comparing the left and right panels in Fig.~\ref{fig:two_FLUKA_realizations}). By applying the average conversion coefficient to the visible energy, we propagate the event-to-event fluctuations to the reconstructed energy.  This leads to finite energy resolution. The rest of this section is devoted to quantifying this effect.

The accuracy of energy reconstruction should depend on the quality of the measurement. Below, we will consider three sets of assumptions about the experiment.
\begin{enumerate}
    \item \emph{CDR thresholds}: Any particle created below the thresholds listed in Table~\ref{tab:threshold} is lost.
    \item \emph{Total charge calorimetry}: Thresholds are set to zero and no information about the hadronic system other than the total ionization charge is used. 
    \item \emph{Detailed event reconstruction}: Thresholds are low and recombination corrections are applied to each particle in the event individually. 
\end{enumerate}
Scenario 1 is motivated by the DUNE CDR \cite{Acciarri:2015uup,Alion:2016uaj}, but with thresholds applied consistently here, to all event stages. We have already presented the average energy budget for it in Fig.~\ref{fig:missing_energy_4GeV}; we now quantify the fluctuations and their impact on energy resolution. Scenario 2 follows the approach advocated in Refs.~\cite{Sorel:2014rka,DeRomeri:2016qwo}, with one improvement: we use separate conversion coefficients for the total charges created by the lepton shower and by the hadronic system, following Ref.~\cite{Grant2018DPFtalk}. Since this uses more information in the event, it can only improve the resolution.

\begin{figure}[t]
	\begin{center}
        \includegraphics[width=\columnwidth]{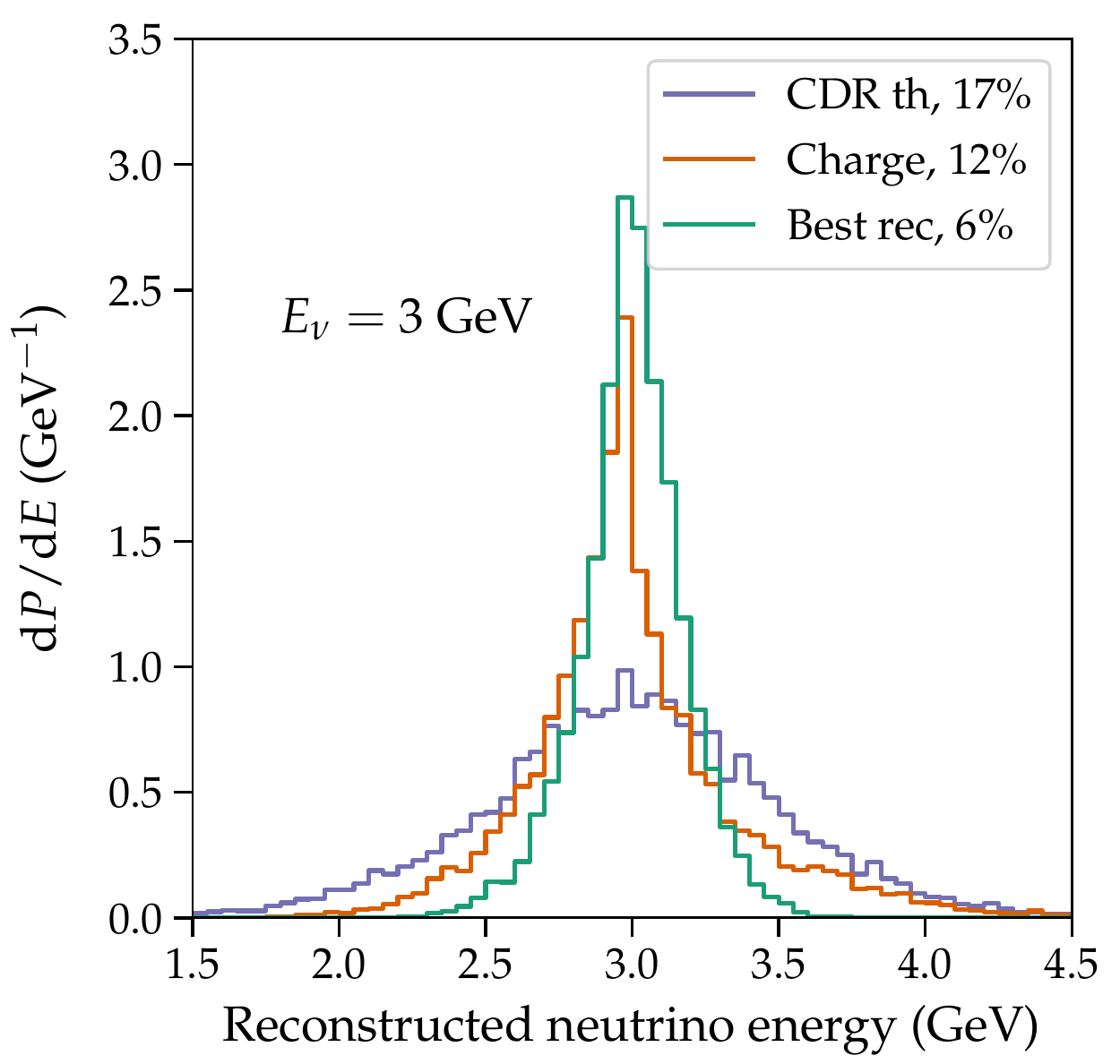}
        \caption{Simulations of reconstructed neutrino energies for $E_\nu=3$ GeV true energy in the CC $\nu_e+^{40}$Ar scattering process. The histograms correspond to three different sets of assumptions, as described in the text.}
        \label{fig:neutrino_resolution}
    \end{center}
\end{figure}

\begin{figure}[h]
	\begin{center}
        \includegraphics[width=\columnwidth]{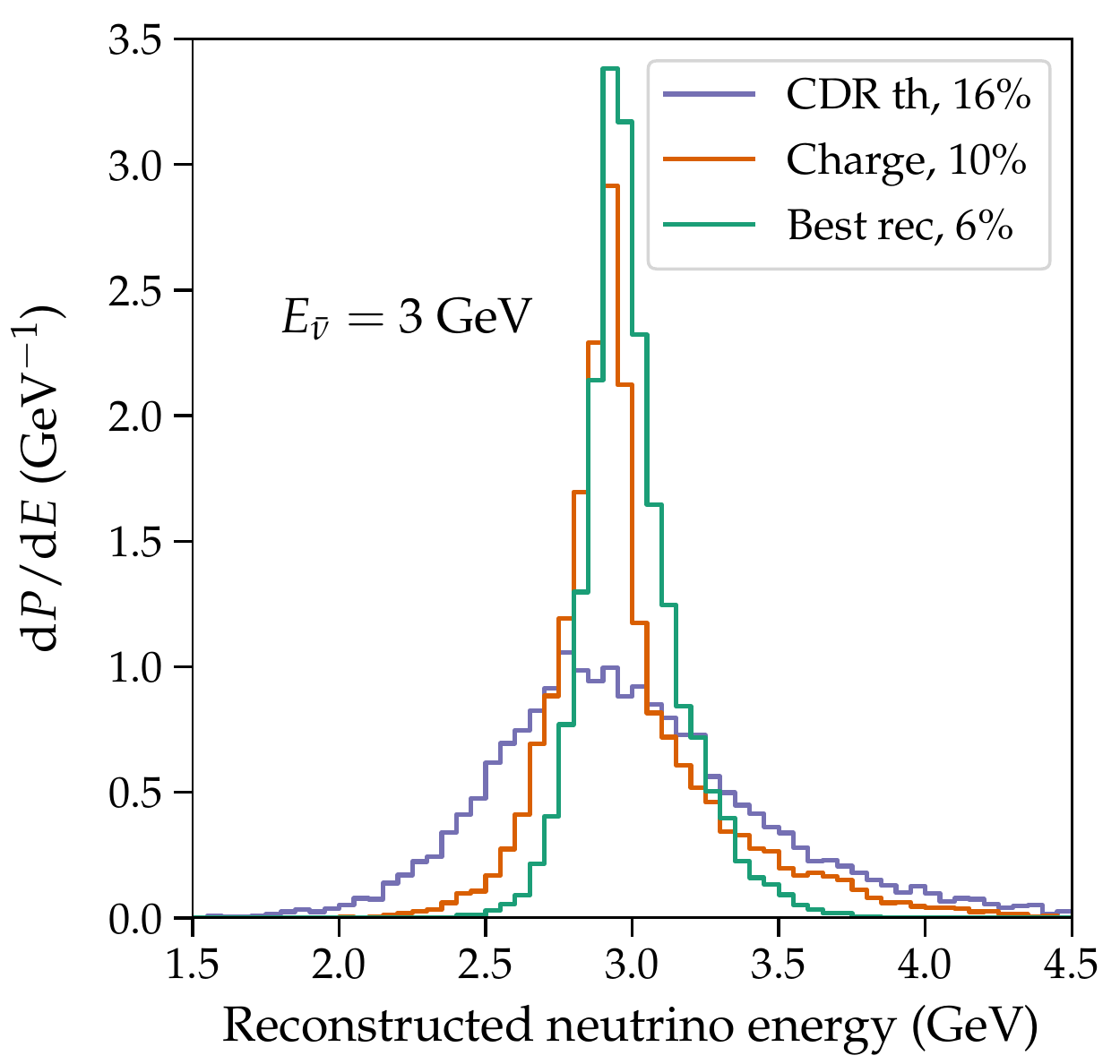}
        \caption{Same as Fig.~\ref{fig:neutrino_resolution}, but for $\bar\nu_e+^{40}$Ar scattering.}
        \label{fig:antineutrino_resolution}
    \end{center}
\end{figure}

\begin{figure*}[t]
	\begin{center}
        \includegraphics[width=\textwidth]{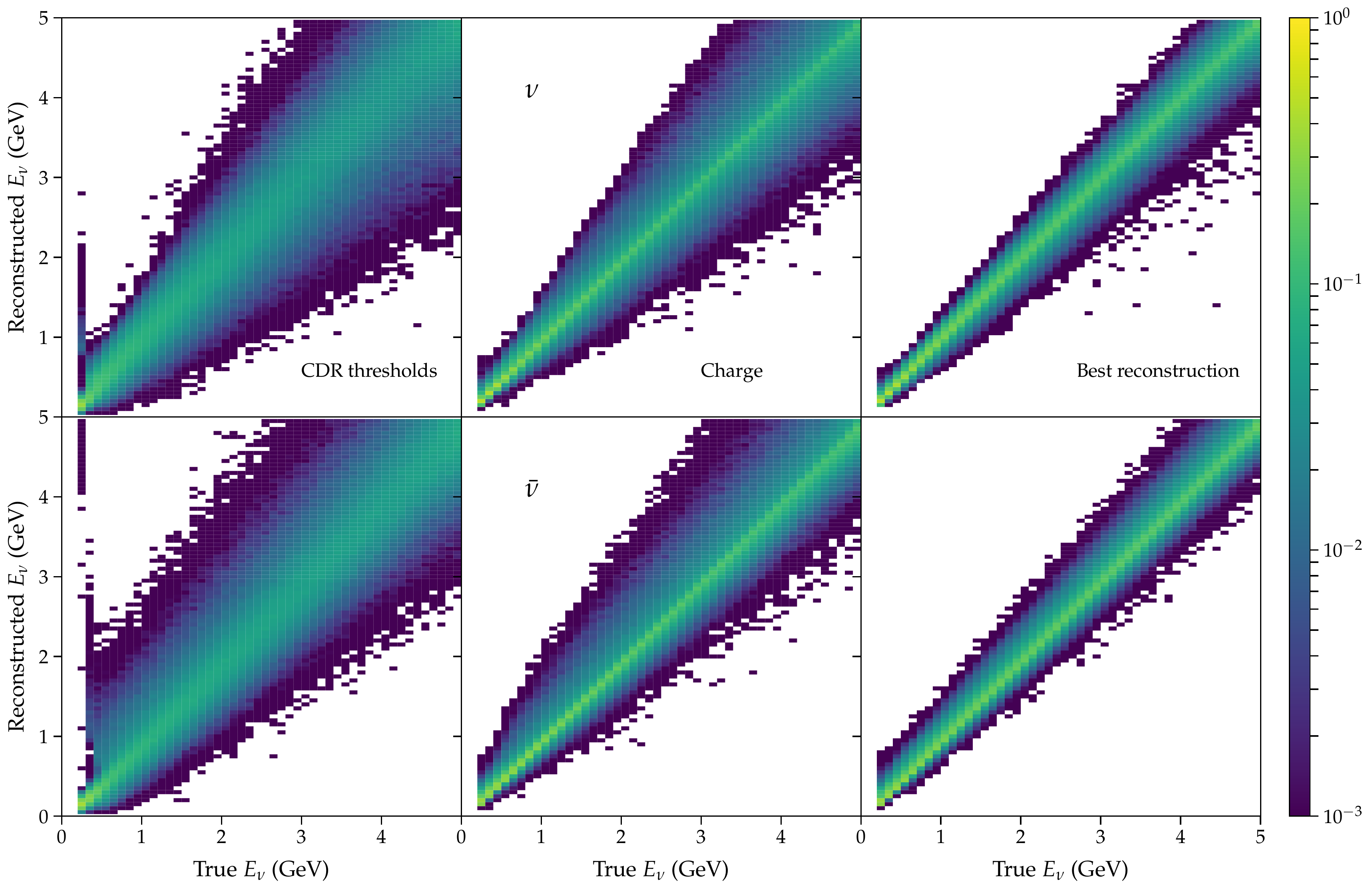}
        \caption{Migration matrices for $\nu_e$ (\textit{top}) and $\bar\nu_e$ (\textit{bottom}) CC scattering in argon in  the three scenarios considered in the text: CDR thresholds (\textit{left}), total charge calorimetry (\textit{middle}), and hit-finding thresholds with charge recombination corrections (\textit{right}).}
        \label{fig:MM_numu}
    \end{center}
\end{figure*}

Finally, the last one is our proposal.  The logic behind it is to quantify how much can be gained by identifying individual particles in DUNE events, while simultaneously lowering thresholds from their CDR values. Specifically, we make the following two key assumptions: 
\begin{itemize}
    \item We posit that event reconstruction in the experiment will allow one to identify all charged particles and use their particle identification information to infer the true energy losses from visible charge, by applying individual recombination corrections. In our simulation, this simply means using the information about true energy losses, rather than ionization charge.
    \item  We lower the thresholds to the values recently demonstrated by the ArgoNeuT collaboration \cite{Acciarri:2018myr}. These low thresholds have already been considered in Fig.~\ref{fig:missing_energy} of the last section.
\end{itemize}
The lowering of the thresholds represents an improvement over scenario 1 above, while the ``un-quenching" procedure improves on scenario 2. To make the comparison more clear-cut, we choose to implement scenario 1 with recombination corrections for particles above the CDR thresholds. The details of the reconstruction procedure are given in Appendix~\ref{appendix:b}.

We simulate energy reconstruction in each of the three scenarios for 10,000 $\nu_e+ ^{40}$Ar events with initial neutrino energy $E_\nu=3$ GeV. The results are depicted in Fig.~\ref{fig:neutrino_resolution}. We see that using the detailed information about events in tandem with low thresholds improves energy resolution. This, by itself, is not surprising, but the degree of improvement is notable. Using particle identification information in scenario 3 to apply individual recombination corrections improves the energy resolution by a factor of two over the total charge method of scenario 2. Lowering the thresholds compared from their CDR values to those of ArgoNeuT results in an improvement of a factor of three (scenario 1 vs scenario 3).

The resolution numbers quoted above are defined in terms of the standard deviations of the corresponding distributions. From Fig.~\ref{fig:neutrino_resolution} we see that the distributions are approximately symmetric around the true energy and one could indeed, to zeroth order, replace them by Gaussians with appropriate widths. That the average reconstructed energy coincides with the input value of 3 GeV should not be at all surprising: we have assumed, by construction, that the energy scale of the experiment is perfectly calibrated. On the other hand, the approximate symmetry of the histograms was not \textit{a priori} guaranteed. Indeed, for comparison, in Fig.~\ref{fig:antineutrino_resolution} we show the result of the same reconstruction simulation, but for antineutrinos of 3~GeV energy. The asymmetry of the green histogram is apparent in this case.  

On closer examination, even in the neutrino case, the shapes of the green or orange histograms do not follow  Gaussian profiles. In particular, notable deviations occur near the true energy value, where the distributions are sharply peaked. Investigation of these deviations leads to an important physical insight, as will be discussed in the next section.

We now turn to the main result of this section, our general simulation of the energy resolution of liquid argon detectors in the 1--5 GeV energy range. We once again consider the three reference scenarios enumerated above: CDR thresholds, total charge calorimetry, and detailed event reconstruction (with charge recombination corrections). In each case, we follow the same procedure already illustrated in the example of the 3 GeV neutrinos, applying it now to a densely spaced grid of true neutrino energy values. The results are depicted in Fig.~\ref{fig:MM_numu}, as a $3\times2$ grid of two-dimensional plots. The columns, left to right, correspond to scenarios 1, 2, and 3.  The rows correspond to neutrinos (top) or antineutrinos (bottom) in the initial state. As we can see, in each case, the procedure yields a two-dimensional relation describing a probability mapping between true and reconstructed neutrinos energies, a so-called migration matrix (MM). Such MMs form the foundation for any oscillation sensitivity studies of a neutrino experiment. 

The simulations are carried out, once again, for electron neutrinos and antineutrinos. It turns out that the case of the $\nu_\mu$ and $\bar\nu_\mu$, once the muon reconstruction procedure is properly defined, leads to numerically very similar answers. We will return to this point in the next section. 

\begin{figure}[t]
	\begin{center}
        \includegraphics[width=\columnwidth]{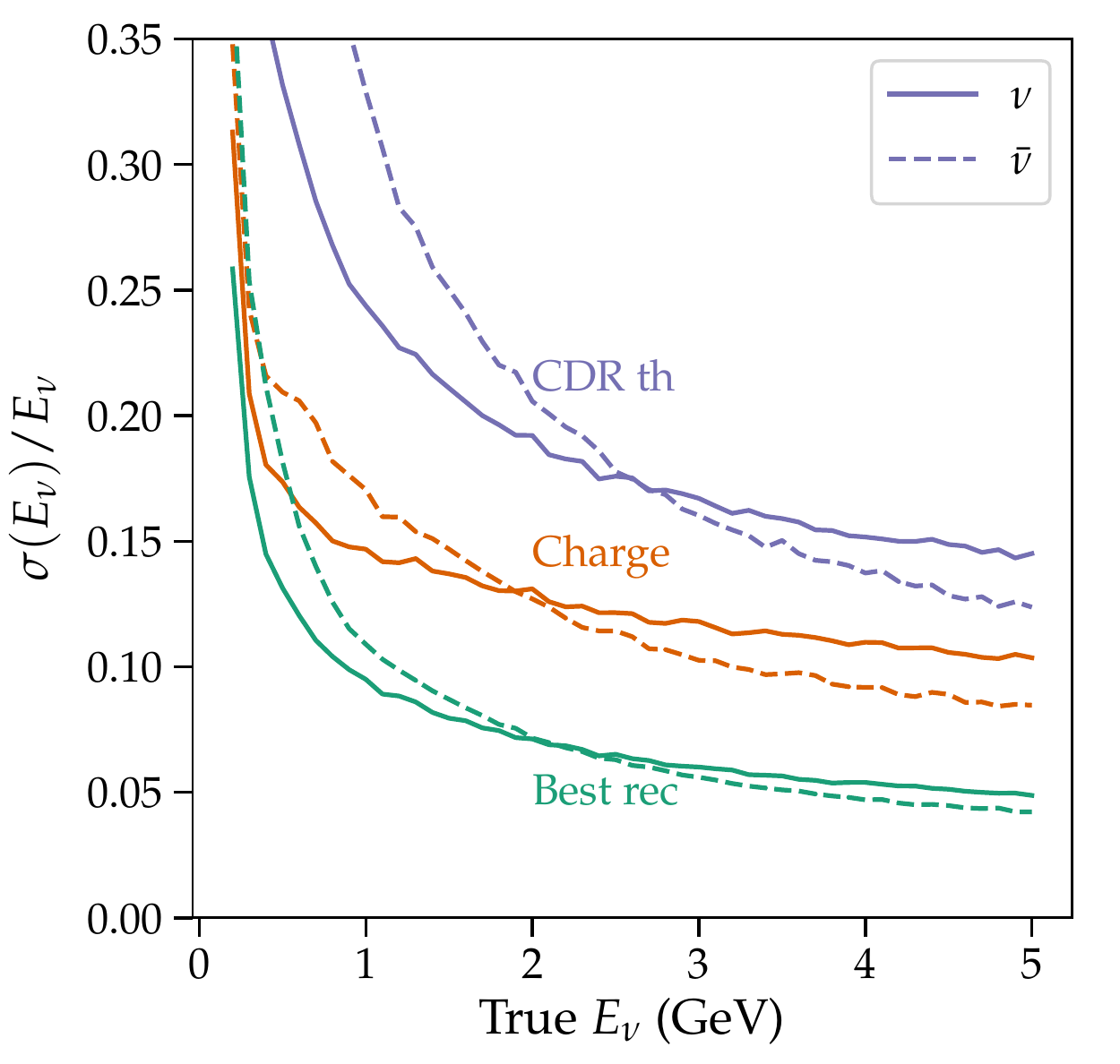}
        \caption{Energy resolution numbers for each of the three reconstruction scenarios considered in the text. Solid curves correspond to $\nu_e$ scattering, and dashed curves to $\bar\nu_e$.}
        \label{fig:resolution}
    \end{center}
\end{figure}

Another relevant observation that will be developed in the next section is the non-Gaussianity of the MMs. We have seen before, in Figs.~\ref{fig:neutrino_resolution} and \ref{fig:antineutrino_resolution}, that at 3~GeV the likelihood distributions in scenarios 2 and 3 are strongly peaked in the center. As is immediately evident in Fig.~\ref{fig:MM_numu}, this behavior persists for all neutrino energies. While the likelihood function is not a Gaussian, it still makes sense to characterize its width by computing the standard deviation of the reconstructed energy distribution, $\sigma(E_\nu)$. Dividing it by $E_\nu$ yields the effective energy resolution of each method. 

Figure~\ref{fig:resolution} shows the results, as a function of the true energy, for neutrinos (solid curves) and antineutrinos (dashed curves). We see that the clear hierarchy of resolutions we already observed in the 3 GeV case persists for all energies.  The only exception is the low-energy window, $E_\nu \lesssim 500$ MeV, where scenarios 2 and 3 yield comparable resolution, both for neutrinos and antineutrinos. Scenario 1 with its high thresholds always performs significantly worse than the other two. 
The relative resolution improves with energy, in each of the three scenarios. As for neutrino vs. antineutrino scattering, the resolution numbers are similar in the range of 2--3 GeV. At higher energies, antineutrinos can be measured more precisely, although the difference is modest.


\section{Discussion}
\label{sect:discussion}

In most general terms, our findings confirm that collecting more information about an event leads to better energy resolution. The extra information can be gained in several ways. Lowering energy thresholds recovers additional particles, most often at the end of showers. Identifying individual particles in showers allows one to apply precise charge recombination corrections to them. In all these cases, resolution degrades as information is discarded. Therefore, it is imperative that all information from the event reconstruction stage is incorporated into energy estimation. 

Beyond this, one of the important benefits of our simulation framework is that it gives a way of estimating quantitatively how much gain one gets from each improvement. This can be used to optimize the strategies for data-taking and analysis at DUNE.  

As the simplest example, let us return to the results of Sec.~\ref{sect:results} and use our scenario 3 there as a reference point. Suppose one had to choose between either (i) raising the thresholds to their CDR values or (ii) using the total ionization of the hadronic system, without recombination corrections. In the first case, we obtain scenario 1 of the previous section, while in the second we get scenario 2. As we found, the resolution in scenario 1 is degraded by nearly a factor of three, changing it from 6\%  to 17\%, while in scenarios 2 and 3 the resolution is degraded only by a factor of two, from 6\% to 12\%. Hence, in this hypothetical situation, one would choose to keep the thresholds low.

We stress that, from the experimental point of view, the low thresholds are certainly not out of reach. Here, one should distinguish trigger thresholds and those of hit-finding. Assuming that data acquisition is triggered by a total visible charge of $\sim$100 MeV, the entire detector should be written out. This would record all charges in the event, including low-energy hadrons and the spray component. To this end, we recall again that the hits from the spray are already seen in ArgoNeuT \cite{Reichenbacher2017talk,Asaadi2017talk,Lepetic2018APStalk,Acciarri:2018myr}, a small surface detector with a level of noise that is higher than what is planned for DUNE. 

A more nuanced question is just how low should the thresholds be. Clearly, going from 0.1 to 50~MeV thresholds is a big step. How high can they be before the impact becomes appreciable? We investigated this question quantitatively. We found that raising thresholds to 0.5 MeV changes the resolution in scenario 3 to 6.5\%, a very modest impact. Setting them to 3 MeV degrades the resolution to 8.2\%. 

The choice of these values is physically motivated. At 0.5 MeV, ArgoNeuT has already demonstrated 50\% efficiency. The value of 3 MeV cuts out the spray component in the event (created in nuclear breakup or by electromagnetic showers). Above that, one starts to significantly cut into charged hadrons. Hence, we conclude that capturing most of the low-energy charged hadrons, and applying the recombination correction to their ionization, causes an improvement in the resolution by a factor of two, from 17\% to 8\%. Including the spray further improves the resolution by 25\%, from 8\% to 6\%.

It should be noted that for charge deposits below 3 MeV one may not be able to do conclusive particle identification. In this case, we appeal to the physics of the process: most of such deposits come from the spray created by low-energy gammas, and thus are electrons. Still, in general, one should distinguish thresholds for particle identification and for simple hit finding.

Returning to the missing energy budget in Fig.~\ref{fig:missing_energy_4GeV}, we see that the CDR thresholds and charge loss to recombination each constitute, on average, the same 20\% fraction of missing hadronic energy \footnote{Here, we are combining losses caused by subthreshold particles created by neutrons, 7\% of $E_{\rm had}$, with those created by other particles, 13\% of $E_{\rm had}$.}. Yet, they affect the resolution differently, as we have seen. This example illustrates a general principle that the impact of a missing energy channel on energy resolution is not directly related to its size in Figs.~\ref{fig:missing_energy_4GeV} and~\ref{fig:missing_energy}. The relevant property is how strongly a given channel fluctuates. While some categories are subject to large fluctuations, others are much more stable.

\begin{figure}[t]
	\begin{center}
        \includegraphics[width=\columnwidth]{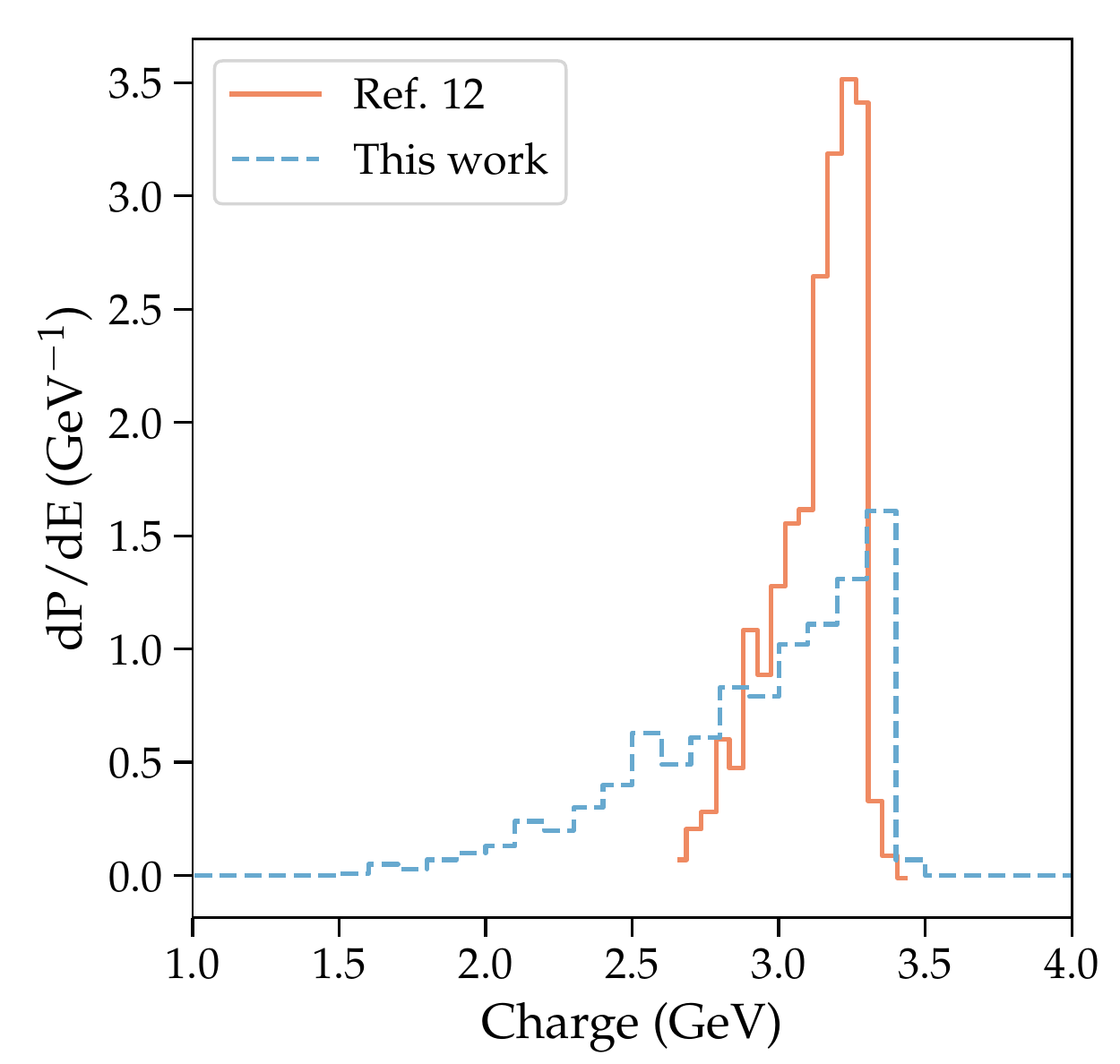}
        \caption{Comparison of the histograms of the total charge created in 4 GeV interactions of $\nu_{e}$: our simulations in scenario 2 vs. the findings of Ref.~\cite{Sorel:2014rka}.}
        \label{fig:comparison}
    \end{center}
\end{figure}

In the former category, we have energy that goes into nuclear breakup. For a 3 GeV neutrino, on average one loses 270 MeV to this category (combined with the small contribution from decay neutrinos), with a standard deviation of 220 MeV. This is the factor limiting the resolution in scenario 3. 

An example of the latter category is provided by electromagnetic showers, which in scenario 3 can be measured with 1.5\% resolution. The $\sim$15\% fraction of the electron energy that disappears in the form of undetectably small charge deposits at the end of the shower is stable ({\it cf.} Fig.\ref{fig:charge_distributions}) and does not have much of an impact on the neutrino energy resolution ({\it cf.} Fig.\ref{fig:missing_energy}). One simply accounts for it by the conversion factor between the visible charge in electromagnetic showers and its true energy. 

Similarly, we found that collecting all the charge from a contained muon track, and accounting for its decay, leads to a 1\% resolution on the muon energy. Thus, both types of leptons can, in principle, be measured well.

One should be cautious when comparing these results to numbers quoted for the charged leptons in other studies. The differences in the details of the assumptions can be very important. For example, some existing simulations consider both exiting and contained muons and quote the average resolution. The resulting resolution is much worse than 1\%, as should be expected, since the energy of exiting muons can be estimated only approximately, from deflection by multiple scatterings. Some simulations, for computational reasons, consider small detector volumes. Such volumes can have a high fraction of exiting muons. Last, but not least, many simulations choose to measure the energy of contained muons by the length of their trajectory, rather than the ionization charge. We modeled this length-based measurement strategy and found that it leads to a muon energy resolution of 4\%, in agreement with the existing literature.

This example allows us to draw two important conclusions. First, one is able to improve the experimental outcome by pruning undesired events, provided the loss of statistics is not overwhelming. Dropping 30\% of escaping muons to obtain a 1\% resolution for the rest may be well justified for studies where energy resolution is a crucial factor (such as mixing angle measurements). We will return to more subtle pruning strategies below. 

Second, since the resolution depends on many factors, it can be illuminating to separate those that are intrinsic to the physics of the shower development from those that can be plausibly overcome in the experiment. In the first category, one can count the fraction of energy that goes into nuclear breakup, which leaves no charge in the detector. In the second category, we count a number of factors, from using track length for muons, to small simulation volumes, to not applying charge recombination corrections, to having high threshold values. Notice that we by no means dismiss the second category; in fact, it may in certain circumstances describe the experimental situation, such as when modeling the near detector, which has a small volume and contamination from cosmics. Yet, we find it helpful to separate such factors from the intrinsic limitations.

\begin{figure}[t]
	\begin{center}
        \includegraphics[width=\columnwidth]{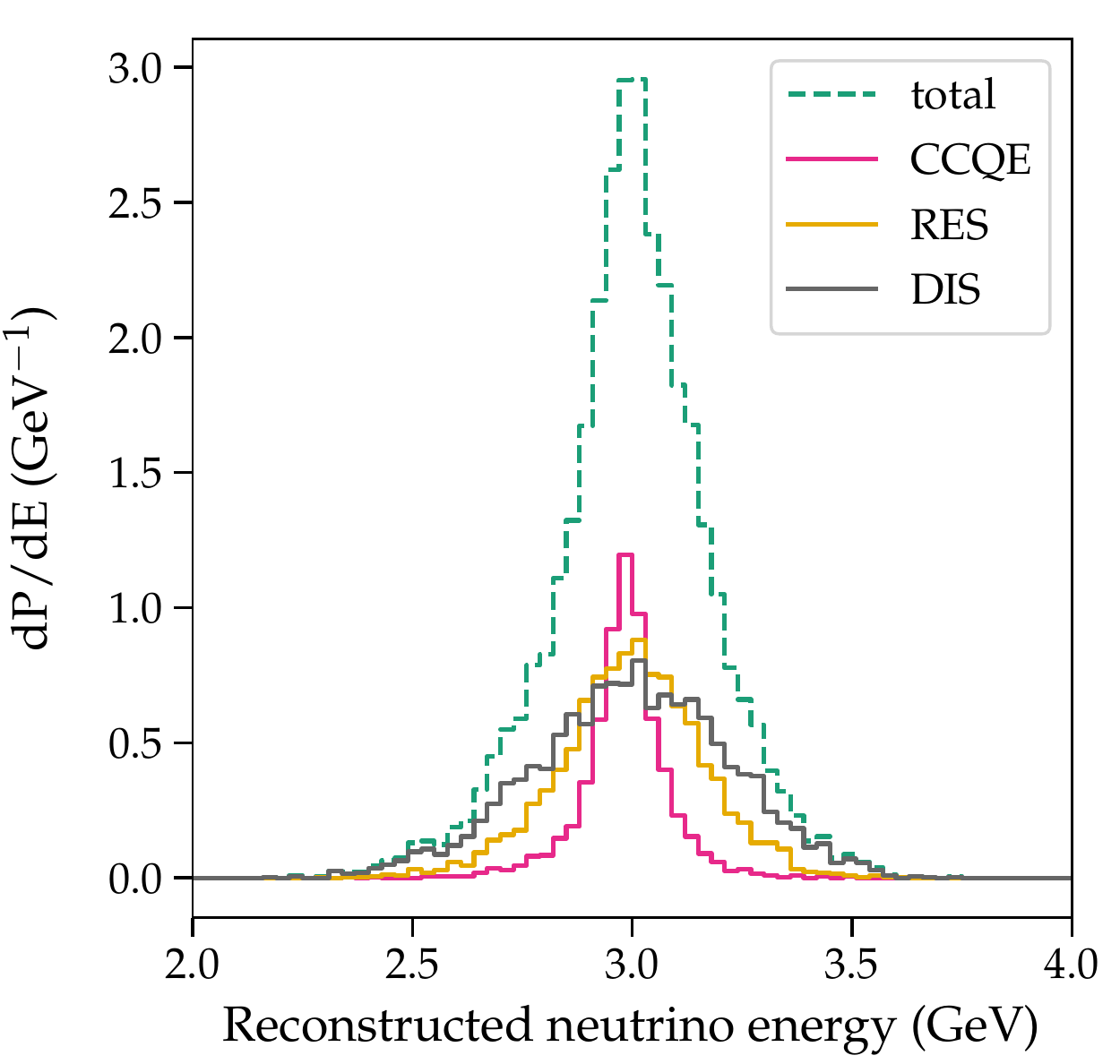}
        \caption{Histogram of reconstructed neutrino energies for $E_\nu=3$ GeV true energy in the CC $\nu_e+^{40}$Ar scattering process in scenario 3, broken down based on event type ({\it cf.} green histogram in Fig.~\ref{fig:neutrino_resolution}).}
        \label{fig:resolution_int}
    \end{center}
\end{figure}

One case where direct literature comparison is possible is with the findings of Ref.~\cite{Sorel:2014rka}, which studied the strategy of measuring the total charge. The corresponding case for us is scenario 2, for which we find a resolution of 12\% at $E_{\nu}=3$ GeV, significantly worse than the 5\% number reported there. To investigate the discrepancies, we compare the histogram of the total ionization charge we find for 4 GeV interactions of $\nu_{e}$ with the corresponding plot in that paper (see the left panel of their Fig.~2). The comparison is presented in Fig.~\ref{fig:comparison}. Both histograms show the same triangular shape, but the one from our simulation is seen to be significantly wider. It would be very desirable to investigate the source of this difference further, as it might provide a good opportunity for model validation between {\tt FLUKA} used by us and {\tt GEANT4} used in {\tt LArSoft} employed in Ref.~\cite{Sorel:2014rka}.

Finally, can one devise an approach that would overcome the 6\% fluctuation we found in nuclear breakup? The answer is yes, at least in principle. Let us return to the green resolution curves in Figs.~\ref{fig:neutrino_resolution} and \ref{fig:antineutrino_resolution}. We noted before that they exhibited non-Gaussian behavior in their central parts, where there were strong peaks around the true neutrino energy. This peaking behavior is physical and can be understood if we separate the events according to their physical process. This is done in Fig.~\ref{fig:resolution_int}. We see that the peak in the center comes from quasi-elastic scattering. This finding, in retrospect, should not be surprising: these events typically have a small fraction of neutrino energy in the hadronic system and the fraction of that energy that goes to nuclear breakup is less. Thus, for some measurements where resolution is paramount it may be reasonable to sacrifice statistics and keep only these events. 

More precisely, to achieve the best physics sensitivity, it will be useful to study oscillations by breaking up the data set into several subsets, according to the amount of hadronic energy and the event topology. This philosophy, is, in fact, already being utilized by the NO$\nu$A experiment, which recently has been breaking up its data into quartiles of hadronic energy~\cite{NOvA:2018gge}.


\section{Conclusions}
\label{sect:conclusions}

Our overarching goal in this paper has been to understand the physics behind the energy resolution of liquid argon detectors. Specifically, we focused on the resolution effects that follow from the non-hermeticity of the detection process. We have identified and characterized individual channels of energy loss and shown how stochastic fluctuations in these channels translate into uncertainties in energy reconstruction. 

Our results are based on a comprehensive simulation campaign, with over 3 million neutrino scattering events fully modeled in the geometry of a single 10 kton module of the DUNE far detector. These simulations are carried out using a pipeline we built expressly for this purpose. The pipeline combines the publicly available tools {\tt GENIE} and {\tt FLUKA} and does not rely on any proprietary tools or information from the DUNE collaboration. Our results should thus be fully reproducible by any interested reader.

While analyzing these results, we found it helpful to classify energy losses into two broad categories: those that can be reduced---or even eliminated---by improving the detector performance and the analysis procedure, and those that are inherent to the detection technique. In the first category, we have particle detection thresholds, which can be lowered with better reconstruction performance. We also have charge recombination, which can be corrected for more accurately if particles on individual tracks are identified. In the second category, we have energy lost to nuclear breakup. For this process, there is no corresponding ionization charge. These losses must be accurately modeled, to get both the energy scale and the resolution right. 

We have quantified how specific assumptions one makes about the loss channels translate into the energy resolution of DUNE. For instance, comparing scenarios 1 and 3 in Sec.~\ref{sect:modelresolution}, we saw that setting thresholds to their CDR values degrades the resolution by as much as a factor of three. At first, this may appear to confirm the findings of Refs.~\cite{Alion:2016uaj} and \cite{DeRomeri:2016qwo}. In fact, any similarity is purely superficial, since our procedures for scenarios 1 and 3 are different from the corresponding approaches in those papers. Indeed, the total charge method of Ref.~\cite{DeRomeri:2016qwo} is represented by our scenario 2, which yields a resolution that is twice as bad as that in scenario 3.

If one could eliminate all losses in the first category, one would be left with losses to nuclear breakup. These losses are caused by multiple scattering of neutrons and, to a lesser extent, also charged hadrons. Fluctuations in this channel give an energy resolution that, in the neutrino case, can be well parametrized by $\Delta E \simeq 10\%/\sqrt{E_{\nu}/{\rm GeV}}$ in the energy range of interest to DUNE. 

As noted in the Discussion, further improvements to the resolution can be possible if one uses a quasi-elastic scattering data subset, which reduces the energy fraction going to nuclear breakup. More generally, the DUNE experiment should benefit from breaking up its data set into subsets of different hadronic energy and event topology.

Our results can have numerous applications:

1. As a source of migration matrices, which serve as the  foundation of any oscillation analysis. By comparing different scenarios, one can establish how the sensitivity of a given oscillation study depends on the experimental performance. In addition to the test cases considered here, one can explore scenarios motivated by specific experimental considerations. For example, one can impose separate thresholds, or detection efficiencies, on particles created by propagating neutrons, since such particles are disconnected from the main event. 

2. As a validation platform for different simulation frameworks. It is noteworthy that our results do not agree with earlier studies. In the case of the DUNE CDR, this should not be surprising, since the Fast Monte Carlo approach used there is very different from the one employed by us: our fully simulated events do not yield themselves to simple Gaussian-smearing prescriptions. The disagreements with the findings of Ref.~\cite{DeRomeri:2016qwo} are, however, notable and should be investigated further, since that study employed {\tt LArSoft} for its simulations.

3. As a motivation for calibration studies. As emphasized repeatedly, our results were obtained using specific versions of {\tt FLUKA} and {\tt GENIE}. To ascertain the accuracy of both missing energy, which affects the energy scale, and the fluctuations, which affect the resolution, one needs to validate the codes using test-beam data. Until recently, the only test-beam measurements for the liquid argon technology were those by LArIAT, the repurposed ArgoNeuT detector with the dimensions 47 cm $\times$  40 cm $\times$ 90 cm, too small to contain charged particles produced by multi-GeV hadrons. Fortunately, very recently high-quality test-beam data have been collected by ProtoDUNE-SP (Single Phase) ~\cite{Abi:2017aow}. These data should go a long way in validating the physics models of particle propagation in liquid argon. Neutron scattering experiments are also very desirable, since neutrons present a special challenge and are responsible for much of the missing energy, as we have seen. To this end, we strongly encourage the Mini-CAPTAIN collaboration~\cite{Berns:2013usa} to release their neutron results~\cite{Jelena2017Praguetalk}. 

4. As a framework for the investigation of cross section uncertainties. It is of great interest to determine the contributions of various such uncertainties to the error budget of DUNE. The present work is a \emph{prerequisite} for this study. This is because cross section errors  propagate through to the oscillation analysis via model-dependent corrections for  missing energy. Indeed, as we saw, the fraction of the neutrino energy that can be captured as ionization depends on the properties of the final-state hadronic system, such as its composition and energy distribution. In turn, those properties are predicted by the model of neutrino-nucleus interactions.

5. As a framework for prioritizing experimental efforts. Clearly, one would like to understand which detector improvements best optimize the neutrino energy scale and resolution. Our results provide the tools to systematically address such questions. For example, one can quantify how much can be gained by increasing neutron efficiency vs. lowering pion thresholds.

6. As points of comparison for the energy resolution studies of other experiments. Besides DUNE, MINOS~\cite{Michael:2008bc}, MINER$\nu$A~\cite{Aliaga:2013uqz}, and NO$\nu$A also employ the calorimetric technique. Our results and methodology, when properly generalized and adapted, could find beneficial applications for them as well. 

Finally, we stress the importance of full detector simulations of DUNE, which would include numerous relevant factors, from electronic noise to cosmogenic and radiogenic backgrounds. We hope that our results can be used as a stepping stone for developing such simulations.


\begin{acknowledgments}
We are grateful to Daniel Cherdack, Richard Gran, Xin Qian, Elizabeth Worcester, and Tingjun Yang, for informative email conversations that sparked our interest and motivated this research. We have greatly benefited from the discussions with Michel Sorel and Ornella Palamara, which followed the presentation of our early results at a Fermilab seminar. We gladly acknowledge conversations with Christopher Mauger, Joshua Klein, and the experimental neutrino team at University of Pennsylvania. We thank Artur Ankowski for valuable comments on the draft and for enlightening physics conversations. We owe special thanks to Paola Sala and the {\tt FLUKA} development team for their support with the package and to Dennis Wright for his answers about {\tt GEANT4}. Finally, we thank the numerous members of the SLAC theory group, whose growing interest in this work over the past year has kept us motivated and excited.  Both authors are supported by the U.S. Department of Energy under Contract No. DE-AC02-76SF00515.
\end{acknowledgments}


\appendix

\section{Composition of the hadronic system}
\label{sect:Appendix_average_composition}

\begin{figure*}[t]
	\begin{center}
        \includegraphics[width=\columnwidth]{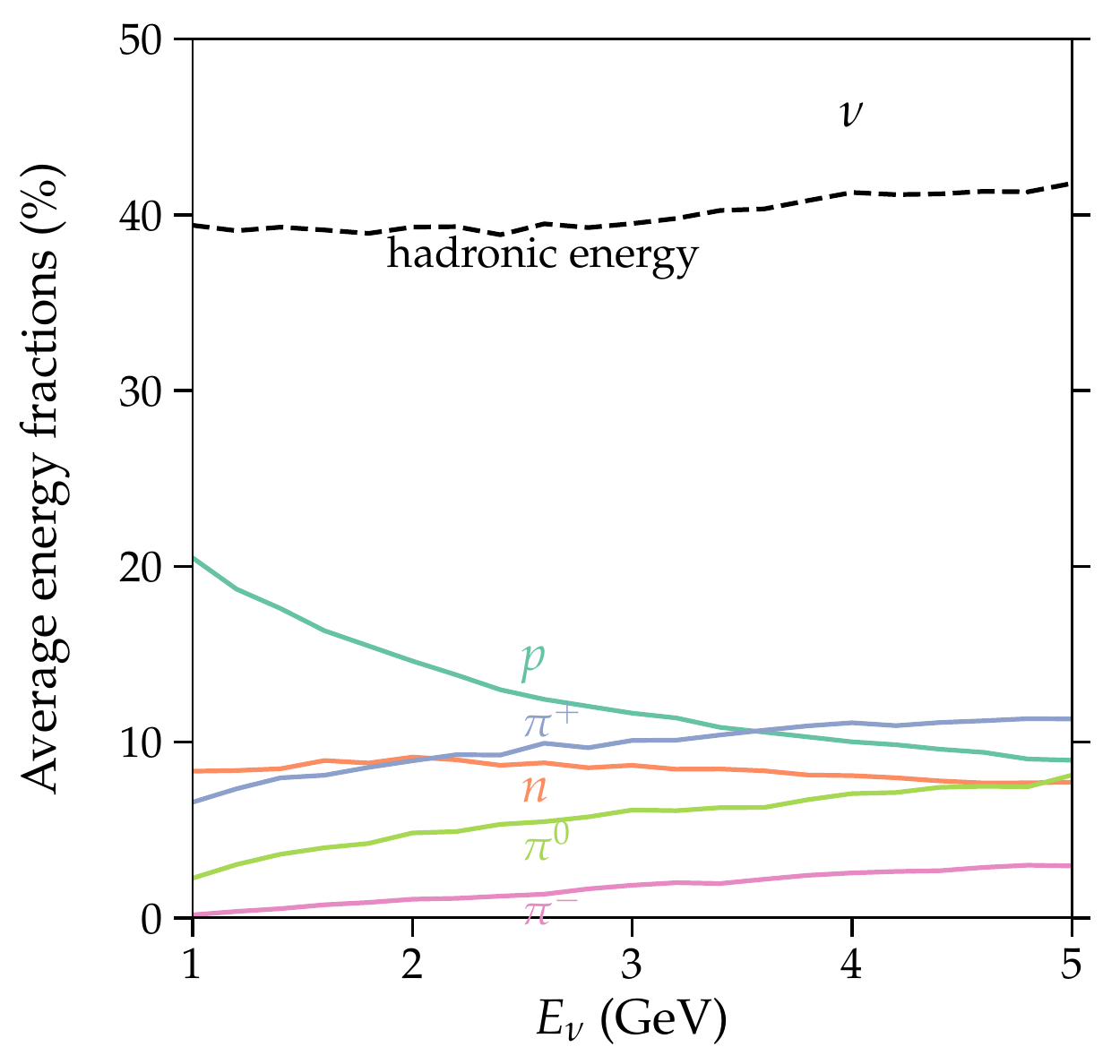}
        \includegraphics[width=\columnwidth]{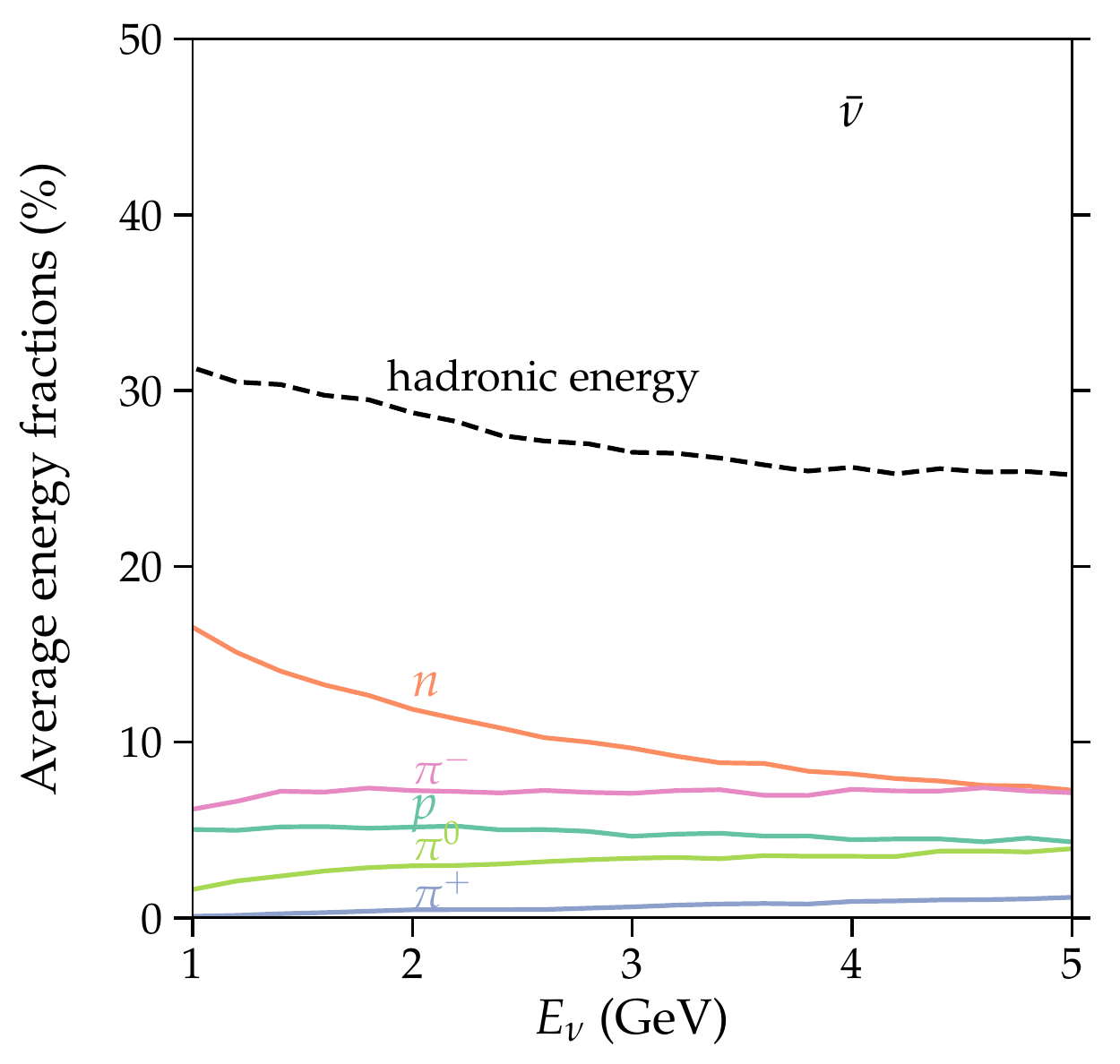}
        \caption{The average energy fractions in different prompt final-state hadrons, for $\nu_e+ ^{40}$Ar (\emph{left panel}) and $\bar\nu_e+ ^{40}$Ar (\emph{right panel}), as a function of the (anti)neutrino energy.}
        \label{fig:had_composion_breakdown}
    \end{center}
\end{figure*}

In Sec.~\ref{sect:primary_interactions}, we have seen that the final-state hadronic system exhibits large event-to-event variations. As illustrated in Fig.~\ref{fig:hadronic_fluctuation}, the variations affect not only the fraction of the total energy that goes into the hadronic system, but also the composition of the system. While some events are dominated by electromagnetic showers (originating from $\pi^0$'s), others have a large fraction of the hadronic energy in charged pions. Still others contain energetic prompt neutrons, which can lead to large energy loss. 

One may wonder how much energy goes into each of these channels on average. We answer this question in Fig.~\ref{fig:had_composion_breakdown}. The results were obtained by averaging over a high-statistics simulation sample and are presented as a function of the incident particle energy, for both neutrino and antineutrino cases.

We see that, while at $E_{\nu}\sim 1$ GeV the hadronic system in $\nu_e+ ^{40}$Ar scattering is dominated by protons, as should be expected from quasi-elastic scattering, at neutrino energies of 3--4 GeV the energy becomes approximately equipartitioned between the various hadron types: $p$, $n$, $\pi^+$, and $\pi^0$. Only negatively charged pions still remain at a lower level, reflecting the fact that the final-state hadronic multiplicity is still moderate, while the overall electric charge of the hadronic system increases by one unit.  The same transition, from neutron-domination to approximate equipartition, is observed for antineutrinos, with the $\pi^+$ component remaining at a lower level in this case, for the same reason as before. 

Once again, we stress that these results apply only \emph{on average}: individual events exhibit large variations and can be dominated by a specific particle type.

\section{Energy reconstruction procedure}
\label{appendix:b}

Here, we briefly describe the procedure used in Sec.~\ref{sect:modelresolution} to find the reconstructed energy in each event. We first separate all charge deposition due to the leading lepton (a muon or an electron) and the hadronic system.  For all charged particles, {\tt FLUKA} records all energy deposited due to ionization. We sum up this deposition throughout the cascade, using the true energy loss for scenarios 1 and 3 and applying the charge recombination formula to find the actual visible charge in scenario 2. Notice that we do not explicitly identify neutrons, only the secondary charged particles (protons and Compton electron spray) that they produce.

We thus obtain the total visible hadronic and leptonic energies, in scenarios 1 and 3, or the corresponding total ionization charges in scenario 2. As the next step, we look up in a database the most likely values of the true energies for the lepton and the hadronic system.  This database is built by simulating a large number of neutrino scattering events and its role is to predict missing energy, given the visible energy / charge.  The leptonic and hadronic energies are summed up at the last step.

Notice that no specific event topology information is used in this procedure. Including it should further improve the resolution.

\newpage

\bibliographystyle{apsrev4-1}
\bibliography{references}

\end{document}